\documentstyle[seceq,epsf,wrapft]{ptptex}

\title{
OPE between the Energy-Momentum Tensor and the Wilson
   Loop in ${\cal N}=4$ Super-Yang-Mills theory
}

\author{
Takehiro {\sc Azuma}\footnote{azuma@gauge.scphys.kyoto-u.ac.jp}
and Hikaru {\sc Kawai}\footnote{hkawai@gauge.scphys.kyoto-u.ac.jp}
}

\inst{
Department of Physics, Kyoto University, Kyoto 606-8502, Japan}

\recdate{
June 11, 2001}

\abst{
  We analyze the operator product expansion $ T_{\mu \nu}(z)
  W[C] $  in ${\cal N}=4$, 4-dimensional Super-Yang-Mills (SYM)
  theory with the $U(N)$ gauge group. We show that a closed Wilson loop
  does not possess an anomalous dimension and that
  only the  shape of the loop is changed by the conformal
  transformation.   
}

\begin{document}
\maketitle

 \section{Introduction}
  Conformal field theory (CFT) is important in modern particle 
 physics in various contexts.  The most powerful 
 tool in the 2-dimensional CFT is the operator product expansion
 (OPE). The OPE between  the energy-momentum tensor and an operator
 extracts its conformal weight.  

 We find it interesting to consider a similar situation regarding the
 Wilson loop in a conformally invariant Yang-Mills theory.  In this
 paper, we investigate the OPE between the energy-momentum 
 tensor and the Wilson loop in ${\cal N}=4$, 4-dimensional
 Super-Yang-Mills (SYM) theory
 with the $U(N)$ gauge group employing dimensional
 analysis and the properties of the energy-momentum
 tensor. We find that the Wilson loop in ${\cal N}=4$ SYM theory
 does not possess an anomalous dimension and that only the deformation
 of the loop occurs under the conformal transformation. 

 Another interesting related topic is the AdS/CFT correspondence.
 \cite{9711200} The  AdS/CFT correspondence
 enables one to evaluate such physical quantities  as
 the multi-point  function \cite{9804058} and the expectation value of 
 the Wilson loop \cite{9803001}$^{,}$\cite{9803002} in the strong coupling
 region. There have
 been ambitious attempts to compute the expectation value of the
 Wilson loop in the strong coupling region by means of quantum field
 theory \cite{0003055}$^{,}$\cite{0010274} for a direct test of the AdS/CFT
 correspondence. In particular, Gross and Drukker
 \cite{0010274} pointed out that the expectation value of a circular 
 Wilson loop in ${\cal N}=4$, 4-dimensional SYM theory is determined by 
 an anomaly in the conformal transformation that relates a circular
 loop to a straight line and computed the expectation value of the
 circular Wilson loop to all orders in the  $\frac{1}{N}$
 expansion.  However, their  analysis is  based on the Feynman gauge, and the
 generalization to the general gauge is non-trivial. We attempt to
 understand the  conformal anomaly through the OPE between the energy-momentum
 tensor and the closed Wilson loop, taking advantage of its gauge
 invariance. 

  This paper is organized as follows.
 Section 2 is devoted to the study of the OPE between $ T_{\mu
 \nu}(z)$ and  $W[C ]$  
      in the $U(N)$  gauge theory by means of 
      dimensional analysis and the properties of the
      energy-momentum tensor. 
 Section 3 presents the computation for the $U(1)$ gauge theory
      as a simple example of the general form investigated in the
      previous  section.
 Section 4 contains the concluding remarks and the outlook for our
         research.
  The appendices contain the proofs of the formulae we derive in full detail.

 \section{General form of the OPE in the ${\cal N}=4$ SYM theory}
   In this section, we develop the OPE between the Wilson loop and the
  energy momentum 
  tensor in the $U(N)$ SYM theory. The bosonic part of
  the Lagrangian and the Wilson loop are as follows:
   \begin{eqnarray}
   {\cal L} &=& \frac{1}{2 G^{2}}   \left[ \frac{1}{2} (F_{\mu
   \nu})^{2} + (D_{\mu} \phi_{i})^{2} + ([\phi_{i}, \phi_{j} ])^{2} +
   \xi (\partial^{\mu} A_{\mu} )^{2}  \right] ,  \nonumber \\ 
  F_{\mu \nu} &=& \partial_{\mu} A_{\nu} - \partial_{\nu}
  A_{\mu} - i  [ A_{\mu}, A_{\nu} ], 
  \label{AZM31nafs} \\
  D_{\mu} \phi_{i} &=& \partial_{\mu} \phi_{i}  - i  [A_{\mu},
  \phi_{i} ],  \label{AZ31cov} \\ 
  W[C ] &=& \frac{1}{N} \textrm{Tr} P \exp \left[ \oint_{C} du 
  \left\{ i A_{\mu}  (y(u))  \frac{dy^{\mu}(u)}{du} + \phi_{i} (y(u)) 
  \theta^{i}(u) \right\}  \right]. \label{AZM31nawilson} 
  \end{eqnarray}
 Throughout this paper, we
 use the following indices: $\mu, \nu, \cdots = 0, 1, 2, 3$ and  $i, j,
 \cdots = 4, 5, \cdots, 9$. Our analysis is
 carried out in Euclidean  space with the metric $g_{\mu \nu} =
 \delta_{\mu \nu} = diag (1, 1,  \cdots, 1 )$. The indices of the scalar
 fields are contracted by $\delta_{ij}$. 
 $G$ denotes the coupling constant.
 $W[ C  ]$ is the  Wilson loop in ${\cal N}=4$, 4-dimensional SYM
      theory, whose derivation is given in detail in a paper of Drukker, Gross
      and Ooguri. \cite{9904191} 
 $y^{\mu}(u)$ represents the coordinates of the Wilson loop. 
      The parameter of the Wilson loop $u$ is an arc length parameter,
      and it satisfies $|\frac{d y_{\mu}(u)}{du}|=1$.
 $\theta_{i}(u)$ is chosen such that $\theta_{i}(u) \theta^{i}(u) = 1$.

 $T_{\mu \nu}(z)$ denotes the energy-momentum tensor of ${\cal N}=4$,
 4-dimensional SYM theory, defined by
 \begin{eqnarray}
   T_{\mu \nu}(z) = \frac{2}{\sqrt{g}} \frac{\delta {\cal L}}{\delta
   g^{\mu \nu}}.
  \end{eqnarray}
  The following two fundamental properties of the energy-momentum
 tensor play a crucial role in our analysis.
   \begin{itemize}
   \item{tracelessness: ${T^{\mu}}_{\mu}(z) = 0$. This implies the
       scale invariance of the action.}
   \item{divergencelessness: $\partial^{\mu} T_{\mu \nu}(z) = 0$. This 
       implies the conservation of the energy and momentum.}
   \end{itemize}  

  \begin{wrapfigure}[8]{r}{6.6cm}
   \epsfxsize = 2.5cm
   \epsfysize = 2.5cm 
     \centerline{\epsfbox{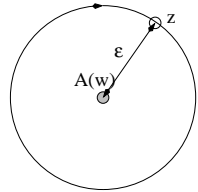}}
    \caption{The contour integral in the 2-dimensional CFT.} 
   \label{2d-wil}
  \end{wrapfigure}
 Before entering the analysis, let us review the well-known OPE
 in the 2-dimensional CFT. In considering the conformal Ward identity,
 we perform a contour integral around the operator $A(w)$. 
 Therefore, ${\cal O}(z-w)^{-1}$ is the order of the weakest singularity that
 contributes to the conformal Ward identity. The OPE is expressed by
  \begin{eqnarray}
   T(z) A(w) &=& \textrm{(lower-dimensional operators)} \nonumber \\
 &+& \frac{h}{(z-w)^{2}} A(w) + \frac{\partial  A(w)}{z-w} + \cdots. 
  \end{eqnarray}
  An example of the lower-dimensional operators is the term of the
  central charge $\frac{\frac{c}{2}}{(z-w)^{4}}$, with 
  $A(w)$ being the energy-momentum tensor.
  The coefficients of $\frac{1}{z-w}$ and $\frac{1}{(z-w)^{2}}$
 represent the translation and the conformal weight of the operator,
 respectively. An important special case is a primary field, on which
  the OPE reduces to
 \begin{eqnarray}
   T(z) A(w) = \frac{h}{(z-w)^{2}} A(w) + \frac{\partial  A(w)}{z-w} +
  \cdots. \label{AZ21toyope} 
 \end{eqnarray}

 \begin{wrapfigure}{r}{6.6cm}
   \epsfxsize = 5cm
   \epsfysize = 3.5cm 
   \centerline{\epsfbox{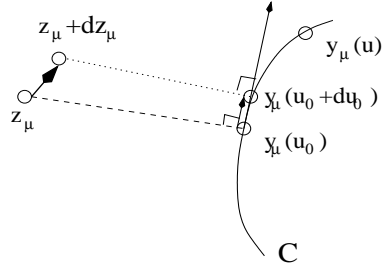}}
    \caption{$y_{\mu}(u_{0})$ is the point nearest to $z_{\mu}$.} 
   \label{curved}
  \end{wrapfigure}
  Let us next consider the OPE between $T_{\mu \nu}(z)$ and  $W[C ]$
  in 4-dimensional Euclidean space. We perform the
  integral over the region wrapping the Wilson loop, and
  this is translated into an integral over the surface of the
  manifold. Let $y^{\mu}(u_{0})$ be the point on the Wilson loop
  nearest to the point $z^{\mu}$. In other words, we take the point
  $y^{\mu}(u_{0})$ so that the vector $z_{\mu} - y_{\mu}(u_{0})$ is
  perpendicular to the tangent vector of the Wilson loop $\frac{
  dy^{\mu}(u_{0})}{du}$:
   \begin{eqnarray}
      (z_{\mu} - y_{\mu}(u_{0})) \frac{d y^{\mu}(u_{0})}{du} = 0.
    \end{eqnarray}
  The dependence of the point $u_{0}$ on the coordinate $z_{\mu}$ is
  given by 
   \begin{eqnarray}
    \frac{\partial u_{0}}{\partial z^{\mu}} = \frac{ \frac{d
    y_{\mu}(u_{0})}{du} }{1 - (z_{\alpha} - y_{\alpha}(u_{0})) \frac{d^{2}
    y^{\alpha}(u_{0})}{du^{2}} }.
    \label{AZ22ds0/dz} 
   \end{eqnarray}
 This can be derived by noting that the vector $(z_{\mu} + dz_{\mu})
 - y_{\mu}(u_{0} + du_{0})$ is also perpendicular to the tangent
 vector $\frac{d  y_{\mu}(u_{0} + du_{0})}{du}$, where $du_{0}$ is the 
  variation of the parameter  $u_{0}$ accompanying an  infinitesimal
 variation of the coordinate $dz_{\mu}$.  
  
 Let $S^{2}(u_{0})$ be the boundary of the 3-dimensional ball
 of a fixed radius $\epsilon$ that is perpendicular to the tangent vector 
 $\frac{d y^{\mu}(u_{0})}{du}$ and has its center  at $y_{\mu}(u_{0})$.
 We wrap the Wilson loop with the surface enveloping these spheres
 $S^{2}(u_{0})$, with  $y^{\mu}(u_{0})$ running  over the whole Wilson loop.
 We define the region inside this enveloping surface as ${\cal
 M}$. Its surface  $\partial {\cal M}$ is, of course, the enveloping surface of
 the spheres $S^{2}(u_{0})$. Utilizing Gauss's theorem, the
 conformal Ward identity for the Wilson loop is 
  \begin{figure}
   \epsfxsize = 7cm
   \epsfysize = 6cm 
   \centerline{\epsfbox{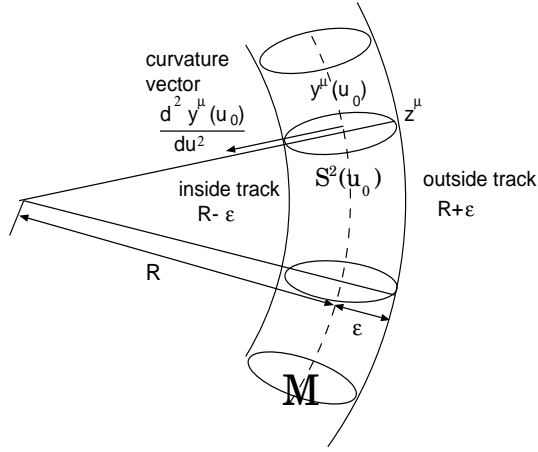}}
    \caption{We perform the integral over the surface 
   enveloping the boundary of the 3-dimensional balls perpendicular to the
   Wilson loop.}
   \label{gausslaw2}
  \end{figure}

  \begin{eqnarray}
  \int_{\cal M} d^{4} z \partial^{\mu} [ T_{\mu \nu}(z) W[C ] 
  v^{\nu}(z) ] = \int du_{0} \int_{S^{2}(u_{0})}  d
  \Omega {\cal C} n^{\mu}  T_{\mu \nu} (z) W[C ] 
  v^{\nu} (z). \nonumber \\ \label{AZ23cwi2}
 \end{eqnarray}
  The meanings of the quantities appearing in the above formula are as
  follows. 
   \begin{itemize}
    \item{ The spacetime integral is performed over the manifold
        ${\cal M}$, in which the Wilson loop is included.}
    \item{ $v^{\nu}(z)$ is a conformal Killing vector. Its explicit
        form is as follows:  
  \begin{eqnarray}
 \hspace{-30mm}   \textrm{Translation:} & & \hspace{3mm}  v^{\nu}(z) =
 \xi^{\nu},  \label{AZM2vtrans} \\
 \hspace{-30mm}   \textrm{Dilatation:}  & & \hspace{3mm} v^{\nu}(z) =
 \lambda z^{\nu}, \label{AZM2vdil} \\
 \hspace{-30mm}   \textrm{Special Conformal Transformation (SCT):} & &
 \hspace{3mm} v^{\nu}(z) = 2 z^{\nu} (b_{\alpha}  z^{\alpha}) - b^{\nu}
 z^{2}. \label{AZM2vsct} 
  \end{eqnarray}   }
   \item{ $d \Omega$ denotes the spherical integral over
  $S^{2}(u_{0})$, and $n^{\mu}$ is the normal vector on the surface
  $\partial {\cal M}$.} 
   \item{The measure ${\cal C}$ results from
  the difference between the inside track and the outside
  track. When the radius of curvature is $R$ and the radius of the
  sphere $S^{2}(u_{0})$ is $\epsilon$, the ratio of the length of
  the  inside 
  track and the outside track is $R-\epsilon : R + \epsilon$.
  Therefore, the measure must be a quantity corresponding to the
  value $1 \mp  \frac{\epsilon}{R}$. 
  Now, the point $z_{\mu}$ resides on the sphere $S^{2}(u_{0})$, and
  the vector $z_{\mu} - y_{\mu}(u_{0})$ corresponds to $\epsilon$. 
  The quantity corresponding to $\frac{1}{R}$ is the
  curvature vector $\frac{d^{2} y_{\mu}(u_{0})}{d u^{2}}$ . Therefore,
  the measure is  
   \begin{eqnarray}
    {\cal C} = 1 - (z_{\alpha} - y_{\alpha}(u_{0})) \frac{d^{2}
    y^{\alpha}(u_{0})}{du^{2}}.  
   \end{eqnarray} 
  The minus sign results from  the fact that the vector $\frac{d^{2}
  y^{\alpha}(u_{0})}{du^{2}}$ is directed at the center of
  curvature.}  
  \end{itemize}
  Since we perform the integral over the sphere $S^{2}(u_{0})$ whose
  surface area is $4 \pi \epsilon^{2}$, the weakest singularity in the 
  OPE that contributes to the conformal Ward identity is ${\cal
  O}(z-y(u_{0}) )^{-2}$.   

  We now investigate the OPE $ T_{\mu \nu}(z) W[C ] $. In 
  the following analysis, we separate the OPE into three parts for
  convenience. 
   \begin{eqnarray}
   \hspace{-10mm}  T_{\mu \nu} (z) W[C ]  = ( T_{\mu \nu} (z)
    W[C ] )_{\textrm{c}} +  ( T_{\mu \nu} (z)
    W[C ] )_{\textrm{vec}} + ( T_{\mu \nu} (z)
    W[C ] )_{\textrm{sca}}.
   \end{eqnarray}
 We hypothesize that the terms corresponding to the lower-dimensional
 operators do not emerge in the OPE $T_{\mu \nu}(z) W[C ]$.
 In this equation, $( T_{\mu \nu} (z) W[C ] )_{\textrm{c}}$ denotes the
       terms containing $W[C ]$ itself  without any insertion of the
       fields into it, which corresponds to the term 
       $\frac{h}{(z-w)^{2}} A(w)$ in the 2-dimensional CFT.
 The other two terms include the insertion of the fields
      $A_{\mu}(y(u_{0}))$ or $\phi_{i}(y(u_{0}))$ into $W[C]$. As we
      find later by means of dimensional analysis,  
      the vector fields and the scalar fields are not inserted
       simultaneously, and we separate the terms into the contribution of the
       vector field and that of the scalar field.
  $( T_{\mu \nu} (z) W[C ] )_{\textrm{vec}}$ and 
  $( T_{\mu \nu} (z) W[C ] )_{\textrm{sca}}$ denote the
       contributions with the insertion of the vector and the scalar
       fields, respectively. 

 \begin{center}
   \begin{tabular}{|c||c|c|c|} \hline
   $T(z) A(w)$ & lower-dimensional operators & $\frac{h}{(z-w)^{2}} A(w)$ &
   $\frac{ \partial A(w)}{z-w}$ \\ \hline 
   $T_{\mu \nu} (z) W[C ]$ &   $-$  & $( T_{\mu \nu} (z) W[C ] )_{\textrm{c}}$
   & $( T_{\mu  \nu} (z) W[C ] )_{\textrm{vec}}$  \\
    & & &  $( T_{\mu \nu} (z) W[C ] )_{\textrm{sca}}$ \\
   \hline  
  \end{tabular}
 \end{center}
  
 \subsection{Contribution of $W[C ]$ itself without field insertion}
  We first investigate the contribution of $W[C ]$ itself $(
  T_{\mu \nu}(z) W[C ] )_{\textrm{c}}$. We express the OPE as a power
  series expansion in $z_{\mu} - y_{\mu}(u_{0})$, where
  $y_{\mu}(u_{0})$ is the nearest point on the Wilson 
  loop to the point $z_{\mu}$. 
  We first list the possible ingredients of this
   contribution:
   \begin{eqnarray}
 & & \left( \frac{dy_{\mu}(u_{0})}{du} \right)^{A}, \hspace{2mm}
  \left( \frac{d^{2} y_{\mu}(u_{0})}{du^{2}} \right)^{B}, \hspace{2mm}
  \left( \frac{d^{3} y_{\mu}(u_{0})}{du^{3}} \right)^{C}, \cdots \nonumber \\
 & & \left( z_{\mu} - y_{\mu}(u_{0}) \right)^{D}, \hspace{2mm}
  \left( \frac{1}{|z-y(u_{0})|} \right)^{E}, \hspace{2mm}
  \left( \theta_{i}(u_{0}) \frac{ d^{2} \theta^{i} (u_{0})}{du^{2}}
  \right)^{F}, \cdots. 
  \end{eqnarray}
 Here we choose the {\it theta} parameter to satisfy $\theta_{i}(u_{0})
  \theta^{i}(u_{0})=1$, and it 
  immediately follows that  $\theta_{i}(u_{0}) \frac{ d \theta^{i}
  (u_{0})}{du} = \frac{1}{2} \frac{d}{du} (\theta_{i}(u_{0})
  \theta^{i}(u_{0})) = 0$.
 The absolute value $|\frac{d
  y_{\mu}(u_{0})}{du} |$  is 1 by definition, and we have $\frac{d 
  y_{\alpha}(u_{0})}{du} \frac{d^{2} y^{\alpha}(u_{0})}{du} =
  \frac{1}{2} \frac{d}{du} (\frac{dy^{\alpha}(u_{0})}{du} \frac{dy_{\alpha} 
  (u_{0})}{du}) = 0$. 
  These powers are restricted by the following conditions.
  \begin{enumerate}
   \item{The singularity that contributes to the conformal Ward
       identity  is at least ${\cal O}(z-y(u_{0}))^{-2}$. Therefore, $D-E
       \leq -2$. }
    \item{The coefficient must have dimensions of
        $(\textrm{length})^{-4}$, and this 
        gives the condition $-B -2C +D - E -2F = -4$.}
    \item{We hypothesize that $z_{\mu} - y_{\mu}(u_{0})$,
  $\theta_{k}(u_{0})$, $\frac{d y_{\mu}(u_{0})}{du}$, $\frac{d^{2} 
  y_{\mu}(u_{0})}{du^{2}}$ and $\frac{d^{3} y_{\mu}(u_{0})}{du^{3}}$ 
  have non-negative powers. Hence $A$, $B$, $C$, $D$ and $F$ 
  must each be 0 or a  positive integer.}
    \item{Since the coefficient must be a tensor of rank two, the
        total number of the indices must be even, so that $A+B+C+D$ is 
        an even number.}
    \item{The result should be invariant under the exchange $u
        \rightarrow -u$, so that $A+3C$ must be an 
        even number.}
  \end{enumerate}
  The second and third conditions lead to the relation $D-E = -4 +
  (B+2C+E+2F) \geq -4$.  Since $B$, $C$, $E$ and $F$ are restricted to
  be 0 or positive integers, the possible singularity in the OPE is thus 
    \begin{eqnarray}
     D-E = -4, -3, -2.
    \end{eqnarray}
 For convenience we classify the contribution of $W[C ]$ itself according to 
 the order of the singularity:
  \begin{eqnarray}
  \hspace{-10mm} ( T_{\mu \nu} (z) W[C ] )_{\textrm{c}} = 
   ( T_{\mu \nu} (z) W[C ] )_{\textrm{c4}} + 
   ( T_{\mu \nu} (z) W[C ] )_{\textrm{c3}} +
   ( T_{\mu \nu} (z) W[C ] )_{\textrm{c2}}. 
  \end{eqnarray}
 Here $( T_{\mu \nu} (z) W[C ] )_{\textrm{c4}}$, $(
 T_{\mu \nu} (z) W[C ] )_{\textrm{c3}}$ and $( T_{\mu \nu} (z) W[C ]
 )_{\textrm{c2}}$ denote the contributions with the singularities of  ${\cal
 O}(z-y(u_{0}))^{-4}$, ${\cal O}(z-y(u_{0}))^{-3}$, and  ${\cal
 O}(z-y(u_{0}))^{-2}$ respectively.
  
 \subsubsection{ Terms with singularities of ${\cal O}(z-y(u_{0}))^{-4}$} 
 We first consider the terms with singularities of  ${\cal
   O}(z-y(u_{0}))^{-4}$.  Since $D-E=-4$, the powers of the other
   ingredients are  
   \begin{eqnarray}
    B=0, \hspace{2mm} C=0, \hspace{2mm} F=0, \hspace{2mm}
    A,D = \textrm{(even)}.
   \end{eqnarray}
 Thus, we find that the possible form of the OPE is 
     \begin{eqnarray}
   ( T_{\mu \nu} (z) W [ C ] )_{\textrm{c4}}  &=&    \frac{1}{24 \pi^{2}
    |z-y(u_{0})|^{4}} \left[ a_{1} g_{\mu \nu} + a_{2} \frac{d
    y_{\mu}(u_{0})}{du} \frac{d y_{\nu}(u_{0})}{du} \right. \nonumber \\
  & & \hspace{-15mm} \left.  + a_{3} \frac{(z_{\mu} - y_{\mu}(u_{0}))
    (z_{\nu} - y_{\nu}(u_{0}))}{|z-y(u_{0})|^{2}}  \right] W[C
    ]. \label{AZM43terms-4}  
   \end{eqnarray}
  The coefficients $a_{1}$, $a_{2}$ and  $a_{3}$ are determined by the
  tracelessness and divergencelessness conditions. The former condition is
  simple, and gives 
   \begin{eqnarray}
   4a_{1} + a_{2} + a_{3} = 0. \label{AZM43trless-4}
   \end{eqnarray}
  The divergencelessness is less simple due to the dependence of the
  point $u_{0}$ on the coordinates, as computed in
  (\ref{AZ22ds0/dz}). The  divergence is given by
   \begin{eqnarray}
   & & \partial^{\mu}  ( T_{\mu \nu} (z) W[C ] )_{\textrm{c4}}
    = \frac{1}{24 \pi^{2}} \left[ - (4a_{1} + 2a_{3}) \frac{z_{\nu} -
   y_{\nu}(u_{0})}{|z-y(u_{0})|^{6}} \right. \nonumber \\
   & & \hspace{5mm} - \left. a_{3} \frac{(z_{\nu} - y_{\nu}(u_{0}))
    [(z_{\alpha} - 
   y_{\alpha}(u_{0})) \frac{d^{2} y^{\alpha}(u)}{du^{2}}  ]}
   { |z-y(u_{0})|^{6} (  1 - (z_{\alpha} - y_{\alpha}(u_{0}))
   \frac{d^{2} y^{\alpha}(u_{0})}{du^{2}} )}  
    + a_{2} \frac{ \frac{d^{2} y_{\nu}(u_{0})}{du^{2}}
   }{|z-y(u_{0})|^{4}}   \right] W[C ]. \nonumber 
   \end{eqnarray}
   We require that only the strongest singularity of ${\cal
   O}(z-y(u_{0}))^{-5}$  
   vanish, because the weaker singularities may be canceled by the
   contribution of the terms in the OPE with weaker singularities. With
   this assumption, the divergencelessness gives the condition  
   \begin{eqnarray}
    4a_{1} + 2a_{3} = 0. \label{AZM43diless-4+}
   \end{eqnarray}

 \subsubsection{ Terms with singularities of  ${\cal O}(z-y(u_{0}))^{-3}$}
   These terms are evaluated in a similar fashion. Since we are now treating
  the terms  of ${\cal O}(z-y(u_{0}))^{-3}$, the powers  must satisfy
  $D-E=-3$, so that  
   \begin{eqnarray}
     B=1, \hspace{2mm} C=0, \hspace{2mm} F=0, \hspace{2mm} A=
     \textrm{(even)}, \hspace{2mm} D= \textrm{(odd)}. 
   \end{eqnarray}
  The possible form of the OPE is thus determined to be
  \begin{eqnarray}
   ( T_{\mu \nu}(z) W[C ] )_{\textrm{c3}}  &=&  \frac{1}{24
  \pi^{2}} \left[ b_{1} \frac{(z_{\mu} - y_{\mu}(u_{0})) (z_{\nu} -
  y_{\nu}(u_{0})) 
  (z_{\alpha} - y_{\alpha}(u_{0})) \frac{d^{2}
  y^{\alpha}(u_{0})}{du^{2}} }{|z-y(u_{0})|^{6}} \right. \nonumber \\
  & & \hspace{-25mm} + b_{2}  \frac{ (z_{\mu} -
  y_{\mu}(u_{0})) \frac{d^{2}  y_{\nu}(u_{0})}{du^{2}}  + (z_{\nu} -
  y_{\nu}(u_{0})) \frac{d^{2} y_{\mu} (u_{0})}{du^{2}} }
  { |z-y(u_{0})|^{4}} \nonumber \\
  & & \hspace{-25mm} + \left. b_{3}  \frac{  \frac{ y_{\mu}(u_{0})}{du}
  \frac{ y_{\nu}(u_{0})}{du} 
  ( z_{\alpha} - y_{\alpha}(u_{0})) \frac{d^{2}
  y^{\alpha}(u_{0})}{du^{2}}  }{|z-y(u_{0})|^{4}} 
  +  b_{4}  g_{\mu \nu} 
  \frac{ (z_{\alpha} - y_{\alpha}(u_{0})) \frac{ d^{2}
  y^{\alpha}(u_{0})}{du^{2}}}{|z-y(u_{0})|^{4}} \right]
  W[C ]. \nonumber \\ \label{AZM43terms-3}  
  \end{eqnarray} 
 These coefficients are again determined by the tracelessness and 
 divergencelessness condition. The former condition is trivial, and yields
  \begin{eqnarray}
   b_{1} + 2 b_{2} + b_{3} + 4 b_{4} = 0. \label{AZM43trless-3}
  \end{eqnarray}
 The latter condition again is less simple, and
 we require only that the strongest singularity ${\cal O}(z-y(u_{0}))^{-4}$
 vanish, together with the results of the previous analysis. This gives
  \begin{eqnarray}
 & & \hspace{-15mm} \partial^{\mu}  [ ( T_{\mu \nu}(z) W[C ] )_{\textrm{c4}} +
  ( T_{\mu \nu}(z) W[C ] )_{\textrm{c3}}  ] \nonumber \\  
  &=& \frac{1}{24 \pi^{2}} \left[ - (a_{3} + b_{1} + 4b_{2} +
   4b_{4}) \frac{ (z_{\nu} - y_{\nu}(u_{0})) [ (z_{\alpha} -
   y_{\alpha}(u_{0})) \frac{d^{2} y^{\alpha}(u_{0})}{du^{2}}  ] }
   { |z-y(u_{0})|^{6} ( 1  - (z_{\alpha} - y_{\alpha}(u_{0}))
   \frac{d^{2} y^{\alpha}(u_{0})}{du^{2}} ) } \right. \nonumber \\
 & & + (b_{4} + a_{2}) \frac{ \frac{d^{2}
   y_{\nu}(u_{0}) }{du^{2}} }{ |z-y(u_{0})|^{4} ( 1- (z_{\alpha} -
   y_{\alpha}(u_{0}))
   \frac{d^{2} y^{\alpha}(s_{0})}{du^{2}} )} \nonumber \\
  & & +   ( - a_{2} - b_{2} + b_{3} -
   b_{4}) \frac{ \frac{d^{2}  y_{\nu}(u_{0})}{du^{2}}  [ (z_{\alpha} -
   y_{\alpha}(u_{0})) \frac{d^{2} y^{\alpha} (u_{0})}{du^{2}}  ]}
   { |z-y(u_{0})|^{4} ( 1 - (z_{\alpha} - y_{\alpha}(u_{0}))
   \frac{d^{2} y^{\alpha}(u_{0}) }{du^{2}} )} \nonumber \\ 
  & &  +  \left. (b_{3} + b_{4}) 
   \frac{ (z_{\alpha} - y_{\alpha}(u_{0})) \frac{d^{2}
   y^{\alpha}(u_{0})}{du^{2}} \frac{d y_{\nu}(u_{0})}{du} }
   { |z-y(u_{0})|^{4} ( 1- (z_{\alpha} - y_{\alpha}(u_{0})) 
   \frac{d^{2}  y^{\alpha}(u_{0})}{du^{2}} ) }  \right] 
   W[C ]. \label{AZM43diless-3} 
  \end{eqnarray}
 This gives the condition  
   \begin{eqnarray}
    a_{3} + b_{1} + 4b_{2} + 4b_{4} = 0, \hspace{3mm} a_{2} + b_{4} =
    0. \label{AZM43diless-3+}
   \end{eqnarray}
  The conditions (\ref{AZM43trless-4}), (\ref{AZM43diless-4+}),
  (\ref{AZM43trless-3}) and (\ref{AZM43diless-3+}) determine the
  coefficients up to two free parameters: 
  \begin{eqnarray}
  & & a_{1} = q, \hspace{2mm} a_{2} = -2q, \hspace{2mm} a_{3} = -2q,
   \nonumber \\
  & & b_{1} = - 2q - 2q', \hspace{2mm} b_{2} = -q  + \frac{q'}{2},
   \hspace{2mm}  b_{3} = - 4q + q', \hspace{2mm} b_{4} = 2q.    
  \end{eqnarray}
  The remaining contribution to the divergence is 
  cancelled together with the terms in the OPE with weaker
  singularities. But we do not pursue their explicit form, because
  they do not affect the  conformal Ward identities.
  
\subsubsection{ Terms with singularities of  ${\cal O}(z-y(u_{0}))^{-2}$}
 We next consider the singularities of ${\cal O}(z-y(u_{0}))^{-2}$. 
 However, it can be shown that these singularities do not contribute
 to the conformal Ward identity using dimensional analysis. In
 performing the spherical integral in the conformal Ward identity
 (\ref{AZ23cwi2}), we utilize the following formulae:
    \begin{eqnarray}
     &(1)& \int_{S^{2}(u_{0})} d \Omega n_{\mu} =  \int_{S^{2}(u_{0})} d
     \Omega n_{\mu} (z_{\nu} - y_{\nu}(u_{0})) (z_{\rho} -
     y_{\rho}(u_{0})) = 0, \label{AZ23int1+} \\
     &(2)& \int_{S^{2}(u_{0})} d \Omega n_{\mu} (z_{\nu} - y_{\nu}(u_{0})) = 
     \frac{4 \pi \epsilon^{3}}{3} \left( g_{\mu \nu} - \frac{d
     y_{\mu}(u_{0})}{du} \frac{d y_{\nu}(u_{0})}{du} \right),
   \label{AZ23int2+} \\
     &(3)& \int_{S^{2}(u_{0})} d \Omega n_{\mu} (z_{\nu} - y_{\nu}(u_{0}))
     (z_{\rho} - y_{\rho}(u_{0})) (z_{\chi} - y_{\chi}(u_{0})) 
     \nonumber \\
     & & \hspace{-12mm}=  \frac{4 \pi \epsilon^{5}}{15} \left[ ( g_{\mu \nu}
     g_{\rho \chi} +  
     g_{\mu \rho} g_{\nu \chi} + g_{\mu \chi} g_{\nu \rho} )  +
     3 \frac{d y_{\mu}(u_{0})}{du} \frac{d y_{\nu}(u_{0})}{du} \frac{d
     y_{\rho}(u_{0})}{du} \frac{d y_{\chi}(u_{0})}{du} \right.
     \nonumber \\
    & &  \hspace{-8mm} -  \left( g_{\mu \nu} \frac{d y_{\rho}(u_{0})}{du}
     \frac{d y_{\chi}(u_{0})}{du}  + g_{\mu \rho} \frac{d
     y_{\nu}(u_{0})}{du} \frac{d y_{\chi}(u_{0})}{du} + g_{\mu \chi}
     \frac{d y_{\nu}(u_{0})}{du} \frac{d y_{\rho}(u_{0})}{du} \right)
     \nonumber \\ 
    & & \hspace{-8mm} -  \left. \left(  g_{\rho \chi}  \frac{d
     y_{\mu}(u_{0})}{du} 
     \frac{d y_{\nu}(u_{0})}{du} +  g_{\nu \chi} \frac{d
     y_{\mu}(u_{0})}{du} \frac{d y_{\rho}(u_{0})}{du} +  g_{\nu \rho}
     \frac{d y_{\mu}(u_{0})}{du} \frac{d y_{\chi}(u_{0})}{du} \right)
     \right]. \nonumber \\
    \label{AZ23int3+} 
   \end{eqnarray}
 Here, $\epsilon$ is the radius of the $S^{2}(u_{0})$ sphere.
 The proof of these formulae is given in full detail in Appendix \ref{sp}. 
 The formula (\ref{AZ23int1+}) indicates that  all we have to do is to
 verify that the power $D$ is an even  number.
 Since ${\cal O}(z-y(u_{0}))^{-2}$ is the order of the weakest singularity
 contributing to the conformal Ward identity, neither the correction
 of the measure ${\cal  
 C}$ nor the positive power of $(z^{\nu} - y^{\nu}(u_{0}))$ in the
 conformal Killing vector contributes any longer. 
 The powers of the ingredients of the OPE must satisfy $B+2C+2F=2$, so
 that   
  \begin{eqnarray}
   (B,C,F) = (2, 0, 0), (0, 1, 0), (0, 0, 1).
  \end{eqnarray}  
 $B$ is thus restricted to be an even number, and we found above 
 that  $A+B+C+D$ and $A+3C$ must be even numbers. It immediately
 follows that $D$ is also an even number, and hence that the terms with 
 singularities of  ${\cal O}(z-y(u_{0}))^{-2}$ do not contribute to
 the conformal Ward identity.

 \subsubsection{The absence of an anomalous dimension  in the
 Wilson loop} 
 We have hitherto derived the contribution of $W[C ]$ itself  to 
 ${\cal O}(z-y(u_{0}))^{3}$, with two parameters $q$ and $q'$. We have 
 
  \begin{eqnarray}
 & & \hspace{-10mm} ( T_{\mu \nu}(z) W[C ] )_{\textrm{c}} =  ( T_{\mu \nu}(z)
 W[C ] )_{\textrm{c4}} + ( T_{\mu \nu}(z) W[C ] )_{\textrm{c3}} 
  +  ( T_{\mu \nu}(z) W[C ] )_{\textrm{c2}} \nonumber \\ 
 &=&   \frac{q}{24 \pi^{2}} \left[ \frac{1}{ |z-y(u_{0})|^{4}}
 \left( g_{\mu \nu}  
  - 2 \frac{d y_{\mu}(u_{0})}{du} \frac{d y_{\nu}(u_{0})}{du}
 \right. \right. 
 \nonumber \\
 & & \left. \left.  - 2 \frac{(z_{\mu} - y_{\mu}(u_{0})) (z_{\nu} -
    y_{\nu}(u_{0}))}{|z-y(u_{0})|^{2}} \right)  \right] W[C ]
 \label{AZ21cnumresult4}  \\ 
 &+& \frac{1}{24 \pi^{2}} \left[  - 2(q+q')  \frac{(z_{\mu} -
 y_{\mu}(u_{0})) (z_{\nu} - y_{\nu}(u_{0})) 
  (z_{\alpha} - y_{\alpha}(u_{0})) \frac{d^{2}
  y^{\alpha}(u_{0})}{du^{2}} }{|z-y(u_{0})|^{6}} \right. \nonumber \\
 & & + \left( - q + \frac{q'}{2} \right)  \frac{ (z_{\mu} -
 y_{\mu}(u_{0})) \frac{d^{2} 
 y_{\nu}(u_{0})}{du^{2}}  + (z_{\nu} -  y_{\nu}(u_{0})) \frac{d^{2}
 y_{\mu} (u_{0})}{du^{2}} }{ |z-y(u_{0})|^{4}}  \nonumber \\  
  & & + (- 4q + q' )   \frac{ \frac{ y_{\mu}(u_{0})}{du}
 \frac{ y_{\nu}(u_{0})}{du} 
  ( z_{\alpha} - y_{\alpha}(u_{0})) \frac{d^{2}
  y^{\alpha}(u_{0})}{du^{2}}  }{|z-y(u_{0})|^{4}} \nonumber \\
  & & + \left. 2q  g_{\mu \nu} 
  \frac{ (z_{\alpha} - y_{\alpha}(u_{0})) \frac{ d^{2}
  y^{\alpha}(u_{0})}{du^{2}}}{|z-y(u_{0})|^{4}}  \right]
  W[C ]  + {\cal O}(z-y(u_{0}))^{-2}. \label{AZ21cnumresult3} 
  \end{eqnarray}
 We now evaluate the conformal Ward identity
 (\ref{AZ23cwi2}) for the contribution of $W[C ]$ itself with respect to the
 translation, dilatation and the special conformal transformation.
 We utilize the formulae (\ref{AZ23int1+}) $-$
 (\ref{AZ23int3+}) in  the spherical integral over $S^{2}(u_{0})$.
 Note the following three points in the computation.
   \begin{itemize}
    \item{Even powers of $z_{\mu} - y_{\mu}(u_{0})$ do not affect 
        the result, as seen from the formula (\ref{AZ23int1+}).}
    \item{The quantity $\frac{d y_{\alpha}(u_{0})}{du} \frac{d^{2}
          y^{\alpha}(u_{0}) }{du^{2}}$ vanishes, because 
          $|\frac{d y_{\alpha}(u_{0})}{du}|=1$.}
     \item{The positive power of $\epsilon$ does not contribute,
         because we set the radius of $S^{2}(u_{0})$ to be a small
         value.} 
   \end{itemize}
  First, we compute the conformal Ward identity for the
  translation, and verify that the translation does not have an
  anomaly:
  \begin{eqnarray}
 & & \hspace{-10mm} \int_{\cal M} d^{4} z \partial^{\mu} [ ( T_{\mu
 \nu}(z) W[C ] )_{\textrm{c}} \xi^{\nu} ] \nonumber \\
 &=& \int du_{0} d \Omega n^{\mu} \left[ 1- (z_{\alpha} -
 y_{\alpha} (u_{0})) \frac{d^{2} y^{\alpha}(u_{0})}{du^{2}}  \right] (
 T_{\mu \nu} (z) W[C ] )_{\textrm{c}} \xi^{\nu} \nonumber \\
 &=& - \int du_{0} d \Omega (z_{\alpha} - y_{\alpha} (u_{0}))
 \frac{d^{2} y^{\alpha}(u_{0})}{du^{2}} n^{\mu} (
 T_{\mu \nu} (z) W[C ] )_{\textrm{c4}} \xi^{\nu} \nonumber \\
 & &  + \int du_{0} d \Omega  n^{\mu} ( T_{\mu \nu} (z) W[C ]
 )_{\textrm{c3}} \xi^{\nu} \nonumber \\
 &=& - \frac{q}{24 \pi^{2} \epsilon^{4}} \frac{4 \pi \epsilon^{3}}{3}
 \int du_{0} \left[ g_{\mu \nu} \left( {g^{\mu}}_{\alpha} - \frac{d
 y^{\mu}(u_{0})}{du} 
 \frac{d y_{\alpha} (u_{0})}{du} \right) \frac{d^{2}
 y^{\alpha}(u_{0})}{du^{2}} \right. \nonumber \\
 & & \hspace{20mm}  - 2 \frac{d y_{\mu} (u_{0})}{du} \frac{d
 y_{\nu}(u_{0})}{du} 
 \left( {g^{\mu}}_{\alpha} - \frac{ d y^{\mu}(u_{0})}{du} \frac{ d
 y_{\alpha}(u_{0})}{du} \right) \frac{d^{2} y^{\alpha}(u_{0})}{du^{2}}
 \nonumber \\
 & & \hspace{20mm}   - 2 \left. \left( g_{\nu \alpha} - \frac{d
 y_{\nu}(u_{0})}{du} \frac{d y_{\alpha} (u_{0})}{du} \right) \frac{d^{2}
 y^{\alpha}(u_{0})}{du^{2}}  \right] \xi^{\nu} W[C ] \nonumber \\ 
 & &  + \frac{1}{24 \pi^{2} \epsilon^{4}} \frac{4 \pi \epsilon^{3}}{3} 
  \int du_{0} \left[ - 2 (q+q') \left( g_{\nu \alpha} - \frac{d
 y_{\nu}(u_{0})}{du} \frac{d  y_{\alpha}(u_{0})}{du} \right) \frac{d^{2}
 y^{\alpha}(u_{0})}{du^{2}} \xi^{\nu} \right. \nonumber \\
 & & \hspace{20mm} + \left( -q + \frac{q'}{2} \right) \left
 ( {g^{\mu}}_{\mu} - \frac{d y^{\mu}(u_{0})}{du} \frac{d
 y_{\mu}(u_{0})}{du} \right) \frac{d^{2} y_{\nu}  
 (u_{0})}{du^{2}} \xi^{\nu} \nonumber \\
 & & \hspace{20mm} + (-4q+q') \frac{d y_{\mu}(u_{0})}{du}
 \frac{d y_{\nu}(u_{0})}{du}  \left( {g^{\mu}}_{\alpha} - \frac{d
 y^{\mu}(u_{0})}{du} \frac{d y_{\alpha}(u_{0})}{du} \right) \frac{d^{2}
 y^{\alpha} (u_{0})}{du^{2}} \xi^{\nu} \nonumber \\
 & & \hspace{20mm} + \left. 2q \left( g_{\nu \alpha} - \frac{d
 y_{\nu}(u_{0})}{du} \frac{d y_{\alpha}(u_{0})}{du} \right) \frac{d^{2}
 y^{\alpha}(u_{0})}{du} \xi^{\nu}  \right] W[C ] \nonumber \\
 &=& - \frac{q}{6 \pi \epsilon} \int du_{0} \frac{d^{2}
 y_{\nu}(u_{0})}{du ^{2}} \xi^{\nu} W[C ] = 0.
  \end{eqnarray}

 We next perform the computation of the conformal Ward identity for
 the dilatation. This computation is similar to that for the
 translation. We separate 
 the conformal Killing vector according to the power of $(z_{\nu} -
 y_{\nu}(u_{0}))$ as 
  \begin{eqnarray}
   v^{\nu}(z) = z^{\nu} = y^{\nu}(u_{0}) + (z^{\nu} -
   y^{\nu}(u_{0})). \label{AZM2vdil+}
  \end{eqnarray}
 We then compute the conformal Ward identity as follows, and we find that, 
 like the translation, the dilatation of the Wilson loop does not
 possess an anomaly.
 \begin{eqnarray}
 & & \int_{\cal M} d^{4} z \partial^{\mu} \left[ ( T_{\mu \nu}(z) W[C ]
 )_{\textrm{c}} \lambda z^{\nu} \right] \nonumber \\
 &=& \int du_{0} d \Omega n^{\mu} [ 1- (z_{\alpha} -
 y_{\alpha} (u_{0})) \frac{d^{2} y^{\alpha}(u_{0})}{du^{2}}  ] (
 T_{\mu \nu} (z) W[C ] )_{\textrm{c}} \lambda [y^{\nu}(u_{0}) +  (z^{\nu} -
 y^{\nu} (u_{0}) )  ] \nonumber \\
 &=& - \int du_{0} d \Omega (z_{\alpha} - y_{\alpha} (u_{0}))
 \frac{d^{2} y^{\alpha}(u_{0})}{du^{2}} n^{\mu} (
 T_{\mu \nu} (z) W[C ] )_{\textrm{c4}} \lambda y^{\nu}(u_{0})  \nonumber \\
 & & + \int du_{0} d \Omega n^{\mu} ( T_{\mu \nu} (z) W[C ]
 )_{\textrm{c4}} \lambda (z^{\nu} - y^{\nu}(u_{0}) ) \nonumber \\ 
 & &  + \int du_{0} d \Omega  n^{\mu} ( T_{\mu \nu} (z) W[C ]
 )_{\textrm{c3}} \lambda y^{\nu}(u_{0}) \nonumber \\
 &=& - \frac{q \lambda}{6 \pi \epsilon} \int du_{0}  \left( 1 +
 \frac{d^{2} y^{\nu} (u_{0})}{du^{2}} y_{\nu}(u_{0}) \right) W[C ]
 \nonumber \\ 
 &=&  - \frac{q \lambda}{6 \pi \epsilon} \int
 du_{0} \left( 1- \frac{d y^{\nu}(u_{0})}{du} \frac{d
 y_{\nu}(u_{0})}{du} \right) W[C ] = 0.  
 \end{eqnarray}
 Finally, we perform the computation of the conformal Ward identity for 
 the special conformal transformation. We again separate the conformal 
 Killing vector in terms of the power of $z^{\nu} - y^{\nu}(u_{0})$ as
 $v^{\nu}(z) = v_{0}^{\nu} + v_{1}^{\nu}(z) + v_{2}^{\nu}(z)$,
 where
  \begin{eqnarray}
   v_{0}^{\nu}    &=& 2 y^{\nu}(u_{0}) (b_{\alpha} y^{\alpha}(u_{0})) -
   b^{\nu} (y(u_{0}))^{2}, \nonumber \\
   v_{1}^{\nu}(z) &=& 2 (z_{\alpha} - y_{\alpha}(u_{0})) (b^{\alpha}
   y^{\nu}(u_{0}) - b^{\nu} y^{\alpha}(u_{0})) + 2 (z^{\nu} -
   y^{\nu}(u_{0})) (b_{\alpha} y^{\alpha}(u_{0})), \nonumber \\
   v_{2}^{\nu}(z) &=& 2 b_{\alpha} (z^{\alpha} - y^{\alpha}(u_{0}))
   (z^{\nu} - y^{\nu}(u_{0})) - b^{\nu}
   (z-y(u_{0}))^{2}. \label{AZM2vsct+} 
  \end{eqnarray}
 The computation of the conformal Ward identity again reveals that the 
 special conformal transformation of the Wilson loop does not possess
 an anomaly:
  \begin{eqnarray}
    & & \hspace{-10mm} \int_{\cal M} d^{4} z \partial^{\mu} [ ( T_{\mu
 \nu}(z) W[C ] 
 )_{\textrm{c}} v^{\nu}(z)  ] \nonumber \\ 
  &=&  \int du_{0} d \Omega n^{\mu} \left[ 1- (z_{\alpha} -
 y_{\alpha} (u_{0})) \frac{d^{2} y^{\alpha}(u_{0})}{du^{2}}  \right] (
 T_{\mu \nu} (z) W[C ] )_{\textrm{c}} v^{\nu}(z) \nonumber \\
 &=& - \int du_{0} d \Omega (z_{\alpha} - y_{\alpha} (u_{0}))
 \frac{d^{2} y^{\alpha}(u_{0})}{du^{2}} n^{\mu} (
 T_{\mu \nu} (z) W[C ] )_{\textrm{c4}} v_{0}^{\nu} \nonumber \\
 & & + \int du_{0} d \Omega n^{\mu} ( T_{\mu \nu} (z) W[C ]
 )_{\textrm{c4}}  v_{1}^{\nu}(z) \nonumber \\ 
 & &  + \int du_{0} d \Omega  n^{\mu} ( T_{\mu \nu} (z) W[C ]
 )_{\textrm{c3}} v_{0}^{\nu} \nonumber \\
 &=& - \frac{q}{6 \pi \epsilon} \int du_{0} \frac{d^{2}
 y_{\nu}(u_{0})}{du^{2}} [  2 y^{\nu}(u_{0}) (b_{\alpha} y^{\alpha}(u_{0})) -
   b^{\nu} (y(u_{0}))^{2} ] W[C ] \nonumber \\
 & &  - \frac{q}{3 \pi \epsilon} \int du_{0} (b_{\alpha}
 y^{\alpha}(u_{0})) W[C ] \nonumber \\
 &=& - \frac{q}{3 \pi \epsilon} \int
 du_{0} \left( 1- \frac{d y^{\nu}(u_{0})}{du} \frac{d
 y_{\nu}(u_{0})}{du} \right)  
 (b_{\alpha} y^{\alpha}(u_{0})) W[C ] = 0.
 \end{eqnarray}

 We have come to the  conclusion that the  Wilson loop possesses no
 anomalous dimension  for the translation, dilatation and
 the special conformal transformation, although the OPE itself has a
 non-trivial form (\ref{AZ21cnumresult4}) and
 (\ref{AZ21cnumresult3}). The Wilson loop is dimensionless 
 in the classical theory,  and this result indicates that the same
 holds true in the quantum theory.   

   \subsection{Contribution of the terms with the field insertion to
 $W[C ]$} 
  In the previous section, we have considered the
  contribution of $W[C ]$ itself. The next step is the analysis of  the
  terms with the insertion of the  fields
  $A_{\mu}(y(u_{0}))$ or $\phi_{i}(y(u_{0}))$ into $W[C ]$. We
  investigate how the fields 
  are inserted into $W[C ]$ by means of dimensional analysis. The vector 
  field $A_{\mu}(y(u_{0}))$ must appear in a gauge invariant way, and 
  we require that the vector field
  contribute in terms of the field strength $F_{\mu
  \nu}(y(u_{0}))$. We also require 
  that the scalar field $\phi_{i}(y(u_{0}))$ be accompanied by
  $\theta^{i}(u_{0})$. We again list the possible ingredients of the
  terms: 
    \begin{eqnarray}
 & & \left( \frac{dy_{\mu}(u_{0})}{du} \right)^{A}, \hspace{2mm} 
  \left( \frac{d^{2} y_{\mu}(u_{0})}{du^{2}} \right)^{B}, \hspace{2mm}
  \left( \frac{d^{3} y_{\mu}(u_{0})}{du^{3}} \right)^{C}, \cdots \nonumber \\ 
 & & (z_{\mu} - y_{\mu}(u_{0}))^{D}, \hspace{2mm}
  \left( \frac{1}{|z-y(u_{0})|} \right)^{E}, \hspace{2mm}
  (F_{\mu \nu}(y(u_{0})))^{G}, \hspace{2mm}
   (D_{\alpha} F_{\mu \nu}(y(u_{0})) )^{H}, \cdots, \nonumber \\
 & &  (\theta^{i}(u_{0}) \phi_{i} (y(u_{0})) )^{I}, \hspace{2mm}
  (\theta^{i}(u_{0}) D_{\mu} \phi_{i} (y(u_{0})) )^{J}, \nonumber \\
 & & \left( \frac{d  \theta^{i}(u_{0})}{du} \phi_{i} (y(u_{0}))
  \right)^{K},  \hspace{2mm}
  (\phi_{i}(y(u_{0})) \phi^{i}(y(u_{0} )) )^{L}, \cdots \nonumber
  \end{eqnarray}
  Dimensional analysis restricts their powers as follows.
   \begin{enumerate}
    \item{Each term must have dimensions of $(\textrm{length})^{-4}$. The
        fields $F_{\mu \nu}(y(u_{0}))$ and
        $\phi_{i}(y(u_{0}))$ have dimensions of
        $\textrm{(length)}^{-2}$, and 
        $\textrm{(length)}^{-1}$, respectively, and this gives the
        condition $-B - 2C + D - E - 2G - 3H - I - 2J - 2K - 2L = -4$.}
     \item{The weakest singularity contributing to the conformal Ward
         identity of the
         Wilson loop is ${\cal O}(z-y(u_{0}))^{-2}$, so that $D-E \leq -2$. }
     \item{The other constraints are almost the same as in the previous
         case:\\ $A, B, C, G, H, I, J$ and  $K$ must be 0 or positive
         integers.} 
     \item{Since the OPE is a tensor of rank two, $A+B+C+D+2G + 3H + J 
         $ is an even number.}
     \item{Invariance under the exchange $u \rightarrow -u$
         requires $A+3C+G+H+K$ to be an even number.}
   \end{enumerate}
  The first two conditions lead to $B + 2C + 2G + 3H + I + 2J +  2K +
  2L \leq  2$. Therefore, the possible powers are
   \begin{eqnarray}
     (B,C,G,H,I,J,K,L) &=& (0,0,1,0,0,0,0,0), \hspace{2mm}
     (0,0,0,0,1,0,0,0), \hspace{2mm}
     (1,0,0,0,1,0,0,0), \nonumber \\
   & &  (0,0,0,0,0,1,0,0),\hspace{2mm} (0,0,0,0,0,0,1,0),\hspace{2mm}
     (0,0,0,0,0,0,0,1), \nonumber \\
   & & (0,0,0,0,2,0,0,0).  \nonumber 
   \end{eqnarray}
  This indicates that the vector fields and the scalar fields are not
  inserted simultaneously, and we consider their contributions
  separately. 

 \subsubsection{Contribution of the Vector Field} 
 In this case, the powers of the fields are
 $(B,C,G,H,I,J,K,L)=(0,0,1,0,0,0,0,0)$ and  $A$ is an odd number. The
 possible form of the terms including the vector field is determined to be
  \begin{eqnarray}
  ( T_{\mu \nu}(z) W[C ] )_{\textrm{vec}} 
 &=&  \frac{1}{N} \textrm{Tr} P w_{L, u_{0}} \frac{i}{4 \pi |z-y(u_{0})|^{3} 
   } \left[ a_{1} \left( (z_{\mu} - y_{\mu}(u_{0})) F_{\nu \alpha}(y(u_{0}))
   \frac{d y^{\alpha} (u_{0})}{du} \right. \right. \nonumber \\
 & & \hspace{-25mm} + \left.  (z_{\nu} - y_{\nu}(u_{0}))
   F_{\mu \alpha}(y(u_{0}))  \frac{d y^{\alpha} (u_{0})}{du}  \right)
   \nonumber \\
 & & \hspace{-25mm} + a_{2} g_{\mu \nu} (z^{\alpha} - y^{\alpha}(u_{0}))
   F_{\alpha \beta}(y(u_{0})) \frac{d y^{\beta}(u_{0})}{du} \nonumber 
   \\
 & & \hspace{-25mm} + a_{3} (z^{\alpha} - y^{\alpha}(u_{0})) 
   \left. \left( F_{\mu
   \alpha} (y(u_{0})) \frac{d y_{\nu}(u_{0})}{du} + F_{\nu \alpha}
   (y(u_{0})) \frac{d y_{\mu}(u_{0})}{du}  \right)  \right] w_{u_{0}, 0},
   \label{AZM32opevec}
  \end{eqnarray}
  where $w_{b,a}$ is the piece of the Wilson loop given by
  \begin{eqnarray}
  w_{b,a} = \exp \left[ \oint^{b}_{a} du \left\{ i A_{\mu}(y(u))  
   \frac{d y^{\mu}(u)}{du} + \phi_{i}(y(u)) \theta^{i}(u) \right\}
   \right], 
   \nonumber 
  \end{eqnarray}
 and $L$ is the length of the Wilson loop. The coefficients
 are determined by the following conditions. First, the tracelessness
 condition of the energy-momentum tensor immediately gives 
   \begin{eqnarray}
    2a_{1} + 4a_{2} - 2a_{3} = 0. \label{AZ2vecconst1}
   \end{eqnarray} 
 In considering the divergence, we again require that the strongest 
 singularity vanish, because the subleading terms may be
 cancelled by the terms in the OPE with weaker singularities:
  \begin{eqnarray}
  & & \partial^{\mu} ( T_{\mu \nu}(z) W[C ] )_{\textrm{vec}}  =
  \frac{1}{N} \textrm{Tr} P 
  w_{L, u_{0}}  \frac{i (a_{1} + a_{2}) }{ 4 \pi |z-y(u_{0})|^{3} 
 } \nonumber \\
  & & \hspace{5mm} \times \left[ -3 \frac{ (z^{\mu} - y^{\mu}(u_{0}))
  (z_{\nu} - y_{\nu}(u_{0})) 
  }{ |z-y(u_{0})|^{2}} F_{\mu \alpha} (y(u_{0})) \frac{dy^{\alpha}
  (u_{0})}{ du} \right. \nonumber \\
  & & \hspace{10mm}  + \left. F_{\nu \alpha} (y(u_{0})) \frac{dy^{\alpha}
  (u_{0})}{du}  \right] w_{u_{0}, 0} + {\cal
  O}(z-y(u_{0}))^{-2}. \nonumber 
 \end{eqnarray}
  The cancellation of the most singular part gives the condition
   \begin{eqnarray}
   a_{1} + a_{2} = 0. \label{AZ2vecconst2}
   \end{eqnarray}
  Finally, we require that this reproduces the conformal Ward identity with 
  respect to the translation. The mere translation of the Wilson loop
  should not have an anomaly. Since we are considering the
  contribution of the vector field, we  
  expect the result to be 
   \begin{eqnarray}
    \int_{\cal M} d^{4} z \partial^{\mu} ( T_{\mu \nu}(z) W[C ]
    )_{\textrm{vec}} \xi^{\nu}  = - \int 
    ds \left( \frac{ \delta W[C ] } 
    { \delta y^{\nu}(s)} \right)_{\textrm{vec}} \xi^{\nu},
    \label{AZ2vecansatz} 
   \end{eqnarray}
 where $( \frac{ \delta W[C ] } { \delta y^{\nu}(s)} )_{\textrm{vec}}$ is
 the variation of the Wilson loop under its deformation, $y_{\mu}(s)
 \to y_{\mu}(s) + \delta y_{\mu}(s)$, with only the vector fields 
 involved.  Here, we do not use the arc length parameter $u_{0}$ but,
 rather,  the general
 parameterization $s$, because the deformation of the Wilson loop
 cannot be defined in the arc length parameter. This is due to the
 fact that the length of the Wilson loop changes under the deformation. The
 relationship between the arc 
 length parameter and the general parameterization is
   \begin{eqnarray}
    \frac{du_{0}}{ds} = |\frac{dy_{\mu}(s)}{ds}|.
   \end{eqnarray}
 Then, the deformation of the Wilson loop is
  \begin{eqnarray}
   \left( \frac{ \delta W[C ] } { \delta y^{\nu}(s)} \right)_{\textrm{vec}} =
   \frac{1}{N} \textrm{Tr} P {\hat w}_{2 \pi, s} i F_{\nu \alpha} (y(s))
   \frac{d y^{\alpha} (s)}{ds}  {\hat w}_{s, 0},
  \label{AZ22deformvec}
  \end{eqnarray}
  where ${\hat w}_{b, a}$ is the piece of the Wilson loop in terms of
  the parameter $s$ defined by
 \begin{eqnarray}
  {\hat w}_{b, a} = \exp \left[ \int^{b}_{a}
  ds \left\{ i A_{\mu}(y(s))  \frac{dy^{\mu}(s)}{ds} +
  |\frac{dy_{\mu}(s)}{ds}| \phi_{i} (y(s)) \theta^{i}(s) \right\}
  \right], 
  \end{eqnarray}
  and the  range of the parameter $s$ is $0 \leq s \leq 2 \pi$.

 Originally, the deformation of the Wilson loop is 
  \begin{eqnarray}
  \left( \frac{\delta W[C ]}{\delta y^{\nu}(s)} \right) 
  &=&   \frac{1}{N} \textrm{Tr} P {\hat w}_{2 \pi, s} \left[ i F_{\nu
  \alpha} (y(s))  \frac{d y^{\alpha} (s)}{ds} +
  |\frac{dy_{\mu}(s)}{ds}|  \theta_{i}(s)  D_{\nu} \phi^{i}( y(s) )
  \right] {\hat w}_{s, 0} \nonumber \\ 
  &-& \frac{d}{ds} \left[ \frac{1}{N} \textrm{Tr} P {\hat w}_{2 \pi, s} 
  \left\{ \phi^{i}(y(s)) 
  \theta_{i}(s) \frac{ \frac{d y_{\nu}(s)}{ds}}{|\frac{d
  y_{\mu}(s)}{ds}| } \right\} {\hat w}_{s, 0} \right],
  \label{AZ2wholedefloop} 
  \end{eqnarray}
  as proven in Appendix \ref{defo}. But we separate the deformation
  into the contribution of vector field  
  and the scalar field for convenience. We refer to the former  terms
  as $( \frac{ \delta W[C ] } { \delta y^{\nu}(s)} )_{\textrm{vec}}$.  

  Computing the conformal Ward identity of the translation for the OPE 
  (\ref{AZM32opevec}), we obtain 
   \begin{eqnarray}
 & &   \int_{\cal M} d^{4}z \partial^{\mu}  ( T_{\mu \nu}(z) W[C ]
    )_{\textrm{vec}}  \xi^{\nu} \nonumber \\
 &=& \frac{1}{N} \int du_{0} \textrm{Tr} P w_{L, u_{0}} \frac{i}{4 \pi
    \epsilon^{3}} \frac{ 4 \pi \epsilon^{3}}{3}  \left[ a_{1} 
    \left\{  \left( {g^{\mu}}_{\mu}  
    - \frac{d y^{\mu}(u_{0})}{du} \frac{d y_{\mu}(u_{0})}{du} \right)
    F_{\nu \alpha} ( y(u_{0}) )  \frac{d y^{\alpha}(u_{0})}{du} \xi^{\nu}
    \right. \right. \nonumber \\
 & & \hspace{5mm}  + \left. \left( {g^{\mu}}_{\nu} - \frac{d
    y^{\mu}(u_{0})}{du} \frac{d y_{\nu}(u_{0})}{du} \right) F_{\mu
    \alpha}( y(u_{0}) ) \frac{d y^{\alpha}(u_{0})}{du} \xi^{\nu}
    \right\} \nonumber \\ 
 & &  + a_{2} \left( {g_{\nu}}^{\alpha} - \frac{d
    y_{\nu}(u_{0})}{du} \frac{d y^{\alpha}(u_{0})}{du} \right) F_{\alpha
    \beta} ( y(u_{0}) ) \frac{d y^{\beta}(u_{0})}{du} \xi^{\nu} \nonumber
    \\
 & & + a_{3} \left. \left( g^{\mu \alpha} - \frac{d
    y^{\mu}(u_{0})}{du} \frac{d y^{\alpha}(u_{0})}{du} \right) 
    \left( F_{\mu \alpha} ( y(u_{0}) ) \frac{d  y_{\nu}(u_{0})}{du} + F_{\nu
    \alpha}( y(u_{0}) ) \frac{d y_{\mu}(u_{0})}{du} \right) \xi^{\nu}
    \right] w_{u_{0}, 0} \nonumber \\ 
 &=& \frac{1}{N} \int du_{0} \textrm{Tr} P w_{L, u_{0}} i \frac{4a_{1} +
    a_{2}}{3}  F_{\nu \alpha}( y(u_{0}) ) \frac{d y^{\alpha}(u_{0})}{du}
    \xi^{\nu} w_{u_{0}, 0}. \label{AZ2vecb}  
  \end{eqnarray}
    Comparing (\ref{AZ2vecb}) with (\ref{AZ2vecansatz}), we obtain
    the constraint
   \begin{eqnarray}
    4a_{1} + a_{2} = -3. \label{AZ2vecconst3}
   \end{eqnarray}
  The coefficients are determined by the three constraints
  (\ref{AZ2vecconst1}),  (\ref{AZ2vecconst2}) and
  (\ref{AZ2vecconst3}) as \begin{eqnarray}
    (a_{1}, a_{2}, a_{3} ) = (-1, 1, 1). 
  \end{eqnarray}
  Similar computations of the conformal Ward identity for the dilatation
  and the special conformal transformation  give only the deformation
  of the Wilson loop. We omit
  the process of the computation, because we have only to replace
  $\xi^{\nu}$ with $y^{\nu}(u_{0})$ and $2y^{\nu}(s) (b_{\alpha}
   y^{\alpha}(s)) - b^{\nu} (y(s))^{2}$ for the dilatation and the
  special conformal transformation, respectively:
  \begin{eqnarray}
    \textrm{(Dilatation)} & & \int_{\cal M} d^{4} z \partial^{\mu} (
   T_{\mu \nu}(z) W[C ] )_{\textrm{vec}} \lambda z^{\nu}  \nonumber \\
  &=& - \int 
   ds \left( \frac{ \delta W[C ]}{\delta y^{\nu}(s)} \right)_{\textrm{vec}}
   \lambda y^{\nu}(s), 
   \label{AZ21vectorboost2} \\
  \textrm{(SCT)} & & \int_{\cal M} d^{4} z \partial^{\mu} (
   T_{\mu \nu}(z) W[C ] )_{\textrm{vec}} (2z^{\nu} (b_{\alpha} z^{\alpha} 
   ) - b^{\nu} z^{2})  \nonumber \\
  &=& - \int ds  \left( \frac{ \delta W[C ]}{\delta
   y^{\nu}(s)} \right)_{\textrm{vec}} [2y^{\nu}(s) (b_{\alpha}
   y^{\alpha}(s)) - b^{\nu} (y(s))^{2}  ].
   \label{AZ21vectorboost3} 
  \end{eqnarray}
 These results correspond to the term $\frac{\partial
 A(w)}{z-w}$ in the OPE of the 2-dimensional CFT (\ref{AZ21toyope}),
 which gives the replacement of the position of $A(w)$ in the
 conformal Ward identity. 

 \subsubsection{Contribution of the Scalar Field}
  We next consider the insertion of the scalar field. In this case,
  the pattern of 
  the possible terms possesses more variety than in the case of the vector
  field. The possible powers are now  
   \begin{eqnarray}
    (B,I,J,K,L) &=& (0,1,0,0,0), (1,1,0,0,0), (0,0,1,0,0), \nonumber \\
   & & (0,0,0,1,0), (0,0,0,0,1),  (0,2,0,0,0),
   \end{eqnarray} 
  and the other powers are $C=G=H=0$. In this case, 
  the singularities of both  ${\cal O}(z-y(u_{0}))^{-3}$ and ${\cal
  O}(z-y(u_{0})))^{-2}$ appear. We separate the terms with the
  insertion of the scalar fields as  
   \begin{eqnarray}
  ( T_{\mu \nu} (z) W[C ] )_{\textrm{sca}} =   ( T_{\mu \nu} (z) W[C ] )_{\textrm{sca3}} +
  ( T_{\mu \nu} (z) W[C ] )_{\textrm{sca2}}. 
   \end{eqnarray}
 Here,  $( T_{\mu \nu} (z) W[C ] )_{\textrm{sca3}}$  and $(
  T_{\mu \nu} (z) W[C ]  )_{\textrm{sca2}}$  denote the terms of
  ${\cal  O}(z-y(u_{0}))^{-3}$ and ${\cal O}(z-y(u_{0})))^{-2}$,
  respectively.  

  We first list the powers of the possible terms  with the
  singularity ${\cal O}(z-y(u_{0}) )^{-3}$.  The vanishing powers are 
  $B=C=G=H=J=K=L=0$. As for 
  the other powers, $I=1$, and $A$, $D$ are even numbers. The possible 
  form of the OPE is thus 
   \begin{eqnarray}
 ( T_{\mu \nu} (z) W[C ] )_{\textrm{sca3}} &=& 
    \frac{1}{N} \textrm{Tr} P w_{L, u_{0}} \frac{\theta_{i}(u_{0})
    \phi^{i}( y(u_{0}) )}{ 24 \pi |z-y(u_{0})|^{3} } \left[ b_{1}
    g_{\mu \nu} + b_{2} \frac{(z_{\mu} - y_{\mu}(u_{0})) (z_{\nu} -
    y_{\nu}(u_{0})) }{|z-y(u_{0}) |^{2}} \right. \nonumber \\ 
 & & \hspace{5mm} + \left.  b_{3} \frac{d y_{\mu}(u_{0})}{du}
    \frac{dy_{\nu}(u_{0}) }{du}  \right] w_{u_{0}, 0}. \label{AZ22scaope3} 
   \end{eqnarray}
  These coefficients are determined by the tracelessness and 
  divergencelessness condition and the conformal Ward identity. The
  tracelessness condition gives 
   \begin{eqnarray}
    4b_{1} + b_{2} + b_{3} = 0. \label{AZ2c1}
   \end{eqnarray}
  We again require that the most singular part of the divergence
  cancel. Taking into account the fact that the derivative operates on 
  the pieces of the Wilson loop $w_{L, u_{0}}$ and $w_{u_{0},
  0}$, we have
   \begin{eqnarray}
   & &  \partial^{\mu} ( T_{\mu \nu} (z) W[C ] )_{\textrm{sca3}} =
     \frac{1}{N} \textrm{Tr} P w_{L, u_{0}} \left[ \frac{\theta_{i}(u_{0})
     \phi^{i} (y(u_{0})) }{ 24 \pi} \left( - \frac{ (3b_{1} + b_{2} )
     (z_{\nu} - y_{\nu}(u_{0})) }{|z-y(u_{0})|^{5}}
     \right. \right. \nonumber \\ 
   & & \hspace{10mm} - b_{2} 
       \frac{ (z_{\alpha} - y_{\alpha}(u_{0})) (z_{\nu} -
     y_{\nu}(u_{0})) \frac{d^{2} y^{\alpha}(u_{0})}{du^{2}}
     }{|z-y(u_{0})|^{5} (1- (z_{\alpha} - y_{\alpha}(u_{0}))
     \frac{d^{2} y^{\alpha}(u_{0})}{du^{2}} )}  \nonumber \\
  & & \hspace{10mm} +  b_{3} 
     \frac{ \frac{d^{2} y_{\nu}(u_{0})}{du^{2}} }{|z-y(u_{0})|^{3} (1-
     (z_{\alpha} - y_{\alpha}(u_{0})) \frac{d^{2}
     y^{\alpha}(u_{0})}{du^{2}} )}  \biggr) \nonumber \\
  & & +  \frac{ \theta_{i} (u_{0}) (D^{\alpha} \phi^{i}
     (y(u_{0})) 
     ) \frac{d y_{\alpha}(u_{0})}{du} \frac{d y_{\nu} (u_{0})}{du} +
     \phi^{i} ( y(u_{0}) ) (\frac{d \theta_{i}(u_{0})}{du})\frac{d y_{\nu}
     (u_{0})}{du}  }
     { 24 \pi |z-y(u_{0})|^{3} (1 - (z_{\alpha} - y_{\alpha}(u_{0}))
     \frac{d^{2} y^{\alpha}(u_{0})}{du^{2}} )} \nonumber \\
   & & \hspace{2mm} \times \left. \left( b_{1} g_{\mu \nu} +
    b_{2} \frac{(z_{\mu} - y_{\mu}(u_{0})) (z_{\nu} - y_{\nu}(u_{0}))
    }{|z-y(u_{0}) |^{2}} + b_{3} \frac{d y_{\mu}(u_{0})}{du}
    \frac{dy_{\nu}(u_{0}) }{du}  \right)  \right] w_{u_{0}, 0}. 
   \label{AZ22scaope3res} 
   \end{eqnarray}
  We require that the most singular part vanish, and we thus have the
  condition  
   \begin{eqnarray}
     3b_{1} + b_{2} = 0. \label{AZ2c2}
   \end{eqnarray}

   The next step is to analyze the terms with singularities of  ${\cal
   O}(z-y(u_{0}))^{-2}$. Here, we list the possible powers of the terms.
   \begin{enumerate}
    \item{$B=I=1$, $J=K=L=0$: In this case, $A$ is an even number
        and $D$ is an odd number.}
    \item{$J=1$, $B=I=K=L=0$: $A$ is an even number and $D$ is an odd
        number.}
    \item{$K=1$, $B=I=J=L=0$: Both $A$ and $D$ are odd numbers.}
    \item{$L=1$, $B=I=J=K=0$: Both $A$ and $D$ are even numbers. This
        term is understood not to contribute to the conformal 
        Ward identity for the same reasoning as in the case of $(
        T_{\mu \nu}(z) W[C ] )_{\textrm{c2}}$. The singularity of this
        contribution is ${\cal O}(z-y(u_{0}))^{-2}$, and only an even 
        power of the tensor $(z_{\mu} - y_{\mu}(u_{0}))$ is
        possible. Therefore, this case is not relevant.}
    \item{$I=2$, $B=J=K=L=0$: Both $A$ and $D$ are even numbers. We
        exclude this contribution for the same reasoning as
        above.}
   \end{enumerate}
  The possible form is thus determined to be
   \begin{eqnarray}
  & &  ( T_{\mu \nu} (z) W[C ] )_{\textrm{sca2}}
  \nonumber \\
  &=&  \frac{1}{N} \textrm{Tr} P w_{L, u_{0}} \left[  
  \frac{ \theta_{i} (u_{0}) \phi^{i}(y(u_{0}))  }{24 \pi
  |z-y(u_{0})|^{3}}  
    \left\{ c_{1} \left( (z_{\mu} - y_{\mu}(u_{0})) \frac{d^{2}
    y_{\nu}(u_{0}) }{du^{2}} +  (z_{\nu} - y_{\nu}(u_{0})) \frac{d^{2}
    y_{\mu}(u_{0}) }{du^{2}}   \right) \right. \right. \nonumber \\
  & & \hspace{5mm} + c_{2} g_{\mu \nu} (z_{\alpha} -
    y_{\alpha} (u_{0})) \frac{d y^{\alpha}(u_{0})}{du^{2}} \nonumber
    \\
  & & \hspace{5mm} + c_{3} \frac{ (z_{\mu} - y_{\mu}(u_{0})) (z_{\nu}
    - y_{\nu} (u_{0})) (z_{\alpha} - y_{\alpha} (u_{0}) ) \frac{d
    y^{\alpha}(u_{0})}{du^{2}} }{|z-y(u_{0})|^{3}} \nonumber \\
  & & \hspace{5mm} + \left. c_{4} \frac{d
    y_{\mu}(u_{0}) }{du} \frac{d y_{\nu}(u_{0})}{du} (z_{\alpha} -
    y_{\alpha} (u_{0})) \frac{d^{2} y^{\alpha}(u_{0})}{du^{2}} 
   \right\}  \nonumber \\
  & & +  \frac{ ( \frac{d \theta_{i}(u_{0})}{du})  \phi^{i}(y(u_{0}))}
    { 24 \pi |z-y(u_{0})|^{3}} c_{5}  \left( (z_{\mu} -
    y_{\mu}(u_{0})) \frac{d 
    y_{\nu} (u_{0})}{du} + (z_{\nu} - y_{\nu} (u_{0})) \frac{d
    y_{\mu}(u_{0})}{ du}  \right) \nonumber \\
  & & + \frac{ \theta_{i}(u_{0})}{24 \pi |z-y(u_{0})|^{3}} 
   \left\{ d_{1} [ (z_{\mu} - y_{\mu}(u_{0})) D_{\nu} \phi^{i} ( y(u_{0}) ) +
    (z_{\nu} - y_{\nu}(u_{0})) D_{\mu} \phi^{i} ( y(u_{0}) )  ] \right.
  \nonumber \\
  & &  \hspace{5mm} + d_{2} g_{\mu \nu} (z_{\alpha} -
  y_{\alpha}(u_{0}) ) D^{\alpha} \phi^{i} ( y(u_{0}) ) \nonumber \\
  & & \hspace{5mm} + d_{3} \frac{ (z_{\mu} - y_{\mu}(u_{0})) (z_{\nu} -
    y_{\nu}(u_{0})) (z_{\alpha} - y_{\alpha}(u_{0})) D^{\alpha}
    \phi^{i}( y(u_{0}) ) }{|z-y(u_{0})|^{2}} \nonumber \\
  & & \hspace{5mm} + d_{4}   \frac{d
    y_{\mu}(u_{0}) }{du} \frac{d y_{\nu}(u_{0})}{du} (z_{\alpha} -
    y_{\alpha} (u_{0})) D^{\alpha} \phi^{i}(y(u_{0})) \nonumber \\
  & & \hspace{5mm} + d_{5} \frac{dy_{\alpha}(u_{0})}{du} (D^{\alpha}
    \phi^{i}( y(u_{0}) ) ) \left( (z_{\mu} - y_{\mu}(u_{0})) \frac{d
    y_{\nu} (u_{0})}{du} + (z_{\nu} - y_{\nu} (u_{0}))
  \frac{d y_{\mu}(u_{0})}{du} \right) \bigr\} \Biggr] w_{u_{0},
  0}. \nonumber \\
 \label{AZ22scaope2}
  \end{eqnarray}
  The tracelessness gives the conditions 
   \begin{eqnarray}
    2c_{1} + 4c_{2} + c_{3} + c_{4} = 0, \hspace{2mm} 
    2d_{1} + 4d_{2} + d_{3} + d_{4} = 0. \label{AZ2c3}
   \end{eqnarray}
  In the analysis of the divergencelessness, we require the strongest
  singularity of ${\cal O}(z-y(u_{0}))^{-3}$ to vanish together with 
  the previous analysis:
  \begin{eqnarray}
 & & \hspace{-12mm}  \partial^{\mu} ( T_{\mu \nu} (z) W[C ] )_{\textrm{sca}}  =
  \frac{1}{N} \textrm{Tr} P w_{L, u_{0}}  
  \left[ \frac{ \theta_{i} (u_{0}) \phi^{i}( y(u_{0}) )}{24 \pi
  |z-y(u_{0})|^{3}}  ( b_{3} + c_{1} + c_{2}) \right. \nonumber \\
  & &  \hspace{-5mm} \times \left( \frac{ \frac{d^{2}
  y_{\nu}(u_{0})}{du^{2}} }{ |z-y(u_{0})|^{3} ( 1- (z_{\alpha} -
  y_{\alpha} (u_{0}))  \frac{d^{2} y^{\alpha}(u_{0})}{du^{2}} )}
  \right. \nonumber \\
  & & \hspace{-5mm} - 3 \left. \frac{ (z_{\alpha} - y_{\alpha}(u_{0}))
  (z_{\nu} - y_{\nu}(u_{0})) 
  \frac{d^{2} y^{\alpha}(u_{0})}{du^{2}} }{ |z-y(u_{0})|^{5}  ( 1-
  (z_{\alpha} - y_{\alpha} (u_{0}))  \frac{d^{2}
  y^{\alpha}(u_{0})}{du^{2}} ) }  \right)  \nonumber \\
  & &  \hspace{-10mm} + \frac{\theta_{i}(u_{0})}{24 \pi} (d_{1} +
  d_{2}) \left( -3 \frac{(z_{\alpha} - y_{\alpha}(u_{0})) (z_{\nu} -
  y_{\nu}(u_{0})) D^{\alpha} \phi^{i}( y(u_{0}) )}{ |z-y(u_{0})|^{5} ( 1-
  (z_{\alpha} - y_{\alpha}(u_{0})) \frac{d y^{\alpha}(u_{0})}{du^{2}}
  ) } \right. \nonumber \\
  & & \hspace{-5mm}  + \left. \left. \frac{ D_{\nu} \phi^{i}(y(u_{0}))
  - \frac{ d y_{\alpha}(u_{0})}{du} \frac{d y_{\nu}(u_{0})}{du}
  D^{\alpha} \phi^{i} ( y(u_{0}) ) }{ |z-y(u_{0})|^{3}  ( 1-
  (z_{\alpha} - y_{\alpha} (u_{0}))  \frac{d^{2}
  y^{\alpha}(u_{0})}{du^{2}} ) }  \right)   \right] w_{u_{0}, 0} 
   + {\cal O}(z-y(u_{0}))^{-2}. \nonumber \\
  \label{AZ22scaope2res}
  \end{eqnarray}
 The divergencelessness thus gives the condition 
  \begin{eqnarray}
   b_{3} + c_{1} + c_{2} = 0, \hspace{3mm} d_{1} + d_{2} = 0. \label{AZ2c4}
  \end{eqnarray}

 We next investigate the conditions to reproduce the conformal Ward
 identity with respect to the translation. Since the translation
 should not have an anomaly, the conformal Ward identity
 is expected to be
  \begin{eqnarray}
 & & \hspace{-15mm}  \int_{\cal M} d^{4} z \partial^{\mu}  \left(
    T_{\mu \nu} (z)  W[C ] \right)_{\textrm{sca}}  
    \xi^{\nu} = - \int ds \left( \frac{ \delta W[C ] }{\delta y^{\nu}
   (s)} \right)_{\textrm{sca}} \xi^{\nu} \nonumber \\
 &=& - \frac{1}{N} \int du_{0} \textrm{Tr} P
   w_{L, u_{0}}    [ \theta_{i}(u_{0}) D^{\nu} \phi^{i}(y(u_{0}))
   \xi_{\nu} ] w_{u_{0}, 0},
  \end{eqnarray}
  where $(\frac{ \delta W[C ] }{\delta y^{\nu} (s)} )_{\textrm{sca}}$ is the 
  scalar field contribution to the deformation of the Wilson loop
  (\ref{AZ2wholedefloop}):
  \begin{eqnarray}
   \left( \frac{ \delta W[C ] }{\delta y^{\nu} (s)} \right)_{\textrm{sca}} &=&
   \frac{1}{N}  \textrm{Tr} P {\hat w}_{2 \pi, s} \left
   [ |\frac{dy_{\mu}(s)}{ds}|\theta_{i}(s) D_{\nu} \phi^{i}( y(s) )
   \right] {\hat w}_{s, 0}  \nonumber \\ 
  &-& \frac{d}{ds} \left[ \frac{1}{N} \textrm{Tr} P {\hat w}_{2 \pi, s} 
   \left\{ \phi^{i} (y(s)) \theta_{i} (s) \frac{ \frac{d y_{\nu}
   (s)}{ds}}{|\frac{d y_{\mu}(s)}{ds}|} \right\}  {\hat w}_{s, 0}
   \right]. \label{AZ22deformsca}  
  \end{eqnarray}
  We compute the conformal Ward identity for the translation utilizing
  the formulae (\ref{AZ23int1+}) $-$ (\ref{AZ23int3+}).
  \begin{eqnarray}
 & &  \hspace{-10mm}  \int_{\cal M} d^{4} z \partial^{\mu}
   ( T_{\mu \nu} (z)  W[C ] )_{\textrm{sca}}  \xi^{\nu} \nonumber \\
 &=& \int du_{0} d \Omega n^{\mu}  \left( 1 - (z_{\alpha} -
   y_{\alpha}(u_{0})) \frac{d^{2} y^{\alpha}(u_{0})}{du^{2}} \right)
   (T_{\mu \nu} (z)  W[C ] )_{\textrm{sca}} \xi^{\nu} \nonumber \\
 &=& -  \int du_{0} d \Omega n^{\mu} (z_{\alpha} -
   y_{\alpha}(u_{0})) \frac{d^{2} y^{\alpha}(u_{0})}{du^{2}} (
   T_{\mu \nu} (z)  W[C ] )_{\textrm{sca3}} \xi^{\nu} \nonumber \\
 & &    +  \int du_{0} d \Omega n^{\nu} (
   T_{\mu \nu} (z)  W[C ] )_{\textrm{sca2}} \xi^{\nu} \nonumber \\
 &=&  \frac{1}{N} \int du \textrm{Tr} P w_{L, u_{0}}
   \left[ \frac{\theta_{i}(u_{0})}{18} \left[ (4d_{1} + d_{2} + d_{3})
   D_{\nu} \phi^{i}(y(u_{0})) \right. \right. \nonumber \\
 & & \hspace{0mm} - \left. (d_{1} + d_{2} + d_{3} + 3d_{5}  )
   \frac{d y_{\alpha}(u_{0}) }{du} \frac{d y_{\nu}(u_{0})}{du}
   D^{\alpha} \phi^{i} ( y(u_{0}) )  \right] \xi^{\nu} \nonumber \\
 & & \hspace{-5mm} + \frac{\theta_{i}(u_{0}) \phi^{i}(y(u_{0}))}{18}
    \left[ (-b_{1} - b_{2} + 4c_{1} + c_{2} + c_{3} ) \frac{ d^{2}
   y_{\nu}(u_{0})}{ du^{2}} \xi^{\nu}  \right] \nonumber \\
 & & + \frac{3c_{5}}{18} \left. \left( \frac{d \theta_{i}( y(u_{0}) )
   }{du} \right)  \phi^{i}( y(u_{0}) ) \frac{d y_{\nu}(u_{0})}{du}
   \xi^{\nu}  \right]  w_{u_{0}, 0} \nonumber \\
 &=& \frac{1}{N} \int du \textrm{Tr} P w_{L, u_{0}}
   \left[ \frac{\theta_{i}(u_{0})}{18} \left( (4d_{1} + d_{2} + d_{3})
   D_{\nu}  \phi^{i}(y(u_{0})) \right. \right. \nonumber \\
 & & \hspace{0mm} - \left. (d_{1} + d_{2} + d_{3} + 3d_{5} + 3c_{5} )
   \frac{d y_{\alpha}(u_{0}) }{du} \frac{d y_{\nu}(u_{0})}{du}
   D^{\alpha} \phi^{i}( y(u_{0}) )  \right) \xi^{\nu} \nonumber \\
 & & \hspace{-5mm} + \frac{\theta_{i}(u_{0}) \phi^{i}(y(u_{0}))}{18}
   \left. \left( (-b_{1} - b_{2} + 4c_{1} + c_{2} + c_{3} - 3c_{5} )
   \frac{ d^{2} y_{\nu}(u_{0})}{ du^{2}} \xi^{\nu} \right)  \right]
   w_{u_{0}, 0}. \nonumber \\ 
  \end{eqnarray}
  This requirement gives the conditions
  \begin{eqnarray}
   & & 4d_{1} + d_{2} + d_{3} = -18, \nonumber \\ 
   & & d_{1} + d_{2} + d_{3} + 3d_{5} + 3c_{5} = 0, \nonumber \\
   & & - 2 b_{3} + 4c_{1} + 4c_{2} + c_{3} - 3c_{5} = 0. \label{AZ2c5}
  \end{eqnarray}

   We have obtained only nine  equations (\ref{AZ2c1}), (\ref{AZ2c2}),
  (\ref{AZ2c3}), (\ref{AZ2c4}) and (\ref{AZ2c5}), while there are 13
  coefficients to  be determined: $b_{1}, \cdots, b_{3}$, $c_{1}, \cdots,
  c_{5}$ and  $d_{1}, \cdots,  d_{5}$. It is impossible to determine
  all the coefficients only by general requirements.

  We next {\it assume} that   the conformal Ward identity for the
  dilatation represents only the deformation of the loop.
  We expect the conformal Ward identity to be 
  \begin{eqnarray}
  & & \hspace{-10mm} \int_{\cal M} d^{4}z \partial^{\mu}
   \left( T_{\mu \nu} (z)  W[C ] \right)_{\textrm{sca}}
   \lambda z^{\nu} 
   = - \int ds \left( \frac{ \delta W[C ]}{\delta y^{\nu}(s)}
   \right)_{\textrm{sca}}  \lambda y^{\nu}(s) \nonumber \\
  & &  \hspace{-10mm} =  \frac{\lambda}{N} \int du_{0} \textrm{Tr} P 
   w_{L, u_{0}}  [ \theta_{i}(u_{0}) D^{\nu} \phi^{i}(y(u_{0}))
   y_{\nu}(u_{0})  + \phi^{i}(y(u_{0})) \theta_{i} (u_{0})  ] w_{u_{0},
   0}.  \label{az23aho1}
  \end{eqnarray}
 Substituting (\ref{AZ22scaope3}) and (\ref{AZ22scaope2})
 into the conformal Ward identity and utilizing the formulae of the
 spherical integration (\ref{AZ23int1+}) $-$ (\ref{AZ23int3+}), we have 
    \begin{eqnarray}
   & &  \hspace{-10mm}  \int_{\cal M} d^{4} z \partial^{\mu}
  [  ( T_{\mu \nu} (z)  W[C ] )_{\textrm{sca}} \lambda z^{\nu}  ] \nonumber \\
  &=& \int du_{0} d \Omega n^{\mu}  \left( 1 - (z_{\alpha} -
   y_{\alpha}(u_{0})) \frac{d^{2} y^{\alpha}(u_{0})}{du^{2}} \right)
  (T_{\mu \nu} (z)  W[C ] )_{\textrm{sca}} \lambda [ y^{\nu}(u_{0}) +
  (z^{\nu} - y^{\nu}(u_{0}) )  ] \nonumber \\
 &=& -  \int du_{0} d \Omega n^{\mu} (z_{\alpha} -
   y_{\alpha}(u_{0})) \frac{d^{2} y^{\alpha}(u_{0})}{du^{2}} (
   T_{\mu \nu} (z)  W[C ] )_{\textrm{sca3}} \lambda y^{\nu}(u_{0}) \nonumber \\
 & &  + \int du_{0} d \Omega n^{\mu} (
   T_{\mu \nu} (z)  W[C ] )_{\textrm{sca3}} \lambda (z^{\nu} - y^{\nu}(u_{0}) )
  \nonumber \\
 & & +  \int du_{0} d \Omega n^{\mu} ( T_{\mu \nu} (z)  W[C ] )_{\textrm{sca2}}
  \lambda y^{\nu}(u_{0}) \nonumber \\  
 &=&  \frac{\lambda}{N} \int du_{0} \textrm{Tr} P
   w_{L, u_{0}} 
   \left[ \frac{\theta_{i}(u_{0})}{18} \left(  (4 d_{1} + d_{2} + d_{3} ) 
   y^{\nu}(u_{0}) D_{\nu} \phi^{i}(y(u_{0})) \right. \right. \nonumber
  \\ 
 & & \hspace{15mm}   - \left. (d_{1} + d_{2} + d_{3} + 3d_{5} )  \frac{d
   y_{\alpha}(u_{0}) }{du} \frac{d y_{\nu}(u_{0})}{du}  D^{\alpha}
   \phi^{i}( y(u_{0}) )  y^{\nu}(u_{0})  \right)
 \nonumber \\
 & & \hspace{10mm} + \frac{\theta_{i}(u_{0}) \phi^{i}(y(u_{0}))}{18}
    \left[ ( - b_{1} - b_{2} + 4c_{1} + c_{2} + c_{3}) \frac{d^{2}
   y^{\alpha} (u_{0}) }{du^{2}} y_{\alpha} (u_{0})  + (3 b_{1} +
  3b_{2} ) \right] \nonumber \\
 & & \hspace{10mm}  + \left. \frac{3 c_{5}}{18} \left( \frac{d
  \theta_{i}(u_{0})}{du} \right) \phi^{i}( y(u_{0}) ) \frac{ d
  y_{\nu}(u_{0})}{du} y^{\nu}(u_{0})  \right] w_{u_{0}, 0} \nonumber \\
 &=& \frac{\lambda}{N} \int du_{0} \textrm{Tr} P  w_{L, u_{0}} 
   \left[ \frac{\theta_{i}(u_{0})}{18} \left(  (4 d_{1} + d_{2} + d_{3} ) 
   y^{\nu}(u_{0}) D_{\nu} \phi^{i}(y(u_{0})) \right. \right. \nonumber \\
 & & \hspace{15mm}   - \left. (d_{1} + d_{2} + d_{3} + 3d_{5} + 3c_{5})
  \frac{d y_{\alpha}(u_{0}) }{du} \frac{d y_{\nu}(u_{0})}{du}  D^{\alpha}
   \phi^{i}( y(u_{0}) )  y^{\nu}(u_{0}) \right)
 \nonumber \\
 & & \hspace{10mm} + \frac{\theta_{i}(u_{0}) \phi^{i}(y(u_{0}))}{18}
    \bigr( ( - b_{1} - b_{2} + 4c_{1} + c_{2} + c_{3} - 3c_{5} )
  \frac{d^{2} y^{\alpha} (u_{0}) }{du^{2}} y_{\alpha} (u_{0})
  \nonumber  \\ 
 & & \hspace{15mm} +  (3 b_{1} + 3b_{2}  - 3c_{5}) 
    \bigr)  \Bigr] w_{u_{0}, 0}.  \label{az23aho2}
 \end{eqnarray}
 Comparing (\ref{az23aho2}) with (\ref{az23aho1}), we obtain the new
 constraint  
  \begin{eqnarray}
    3b_{1} + 3b_{2} - 3c_{5} = -18.
  \end{eqnarray}

 We still have not obtained sufficient constraints on the
 coefficients. However, we can verify that, once we assume that the
 Wilson loop undergoes only a deformation for the dilatation,
 the same is true for the special conformal transformation. By imposing
 all the constraints, the coefficients can be  expressed in terms of
 the three free parameters $\alpha$, $\beta$ and $\gamma$ as
  \begin{eqnarray}
   & & b_{1} = 2 - \frac{\alpha}{10} + \frac{\beta}{2} , \hspace{2mm}
       b_{2} = -6 +  \frac{3 \alpha}{10} + \frac{3 \beta}{2} ,
       \hspace{2mm}
       b_{3} = - 2 + \frac{\alpha}{10} - \frac{\beta}{2}, \nonumber \\
   & & c_{1} = 1 +  \frac{\alpha}{10} - \frac{\beta}{2} +
       \frac{\gamma}{5}, \hspace{2mm}
       c_{2} = 1 -  \frac{\alpha}{5} + \beta -  \frac{\gamma}{5},
       \hspace{2mm} 
       c_{3} = -3 + \frac{3 \alpha}{5} - 3 \beta - \frac{3 \gamma}{5}, 
       \nonumber \\
   & & c_{4} = -3 + \gamma, \hspace{2mm}
       c_{5} = 2 + \frac{\alpha}{5} - \beta, \nonumber \\
   & & d_{1} = -4 + \frac{\alpha}{5}, \hspace{2mm}
       d_{2} = 4 - \frac{\alpha}{5},  \hspace{2mm}
       d_{3} = -6 - \frac{3 \alpha}{5}, \hspace{2mm}
       d_{4} = -2 + \alpha, \hspace{2mm}
       d_{5} = \beta.
 \end{eqnarray}
 The conformal Ward identity
 with respect to the special conformal transformation is computed to be 
  \begin{eqnarray}
 & & \hspace{-10mm}  \int_{\cal M} d^{4}z \partial^{\mu}
  [ ( T_{\mu \nu}(z) W[C ] )_{\textrm{sca}}
   ( 2z^{\nu} (b_{\alpha} z^{\alpha}) - b^{\nu} z^{2})  ] \nonumber \\
 &=& - \int ds \left( \frac{ \delta W[C ] }{\delta y^{\nu} (s)
   } \right)_{\textrm{sca}} [ 2 y^{\nu}(s) (b_{\alpha} y^{\alpha}(s)) -
   b^{\nu} (y(s))^{2}  ],
  \end{eqnarray}
 independent of the values of $\alpha$, $\beta$ and $\gamma$.

 \subsection{Summary of the result}
  We have investigated the general form of the OPE $T_{\mu
  \nu}(z) W[C ] $ by means of dimensional analysis, the scale
  invariance of the theory and the conservation laws of the energy and 
  the momentum.  We have imposed the condition 
  that the conformal Ward identity for the translation  represent a  
  mere shift of the loop, which  is, per
  se, a natural physical requirement. We have further {\it
  assumed} that the same is true of the {\it dilatation}.  Then, the
  general form of the OPE is  determined up to five free parameters: 
  \begin{eqnarray}
 & &    T_{\mu \nu}(z) W[C ]  = 
  ( T_{\mu \nu}(z) W[C ] )_{\textrm{c}} + ( T_{\mu \nu}(z)
  W[C ] )_{\textrm{vec}} + ( T_{\mu \nu}(z) W[C ] )_{\textrm{sca}}
  \nonumber \\
 &=&   \frac{q}{24 \pi^{2}} \left[ \frac{1}{ |z-y(u_{0})|^{4}}
 \left( g_{\mu \nu}  
  - 2 \frac{d y_{\mu}(u_{0})}{du} \frac{d y_{\nu}(u_{0})}{du}
  - 2 \frac{(z_{\mu} - y_{\mu}(u_{0})) (z_{\nu} -
    y_{\nu}(u_{0}))}{|z-y(u_{0})|^{2}} \right)  \right] W[C ]
  \nonumber \\ 
 &+& \frac{1}{24 \pi^{2}} \left[  - 2(q+q')  \frac{(z_{\mu} -
 y_{\mu}(u_{0})) (z_{\nu} - y_{\nu}(u_{0})) 
  (z_{\alpha} - y_{\alpha}(u_{0})) \frac{d^{2}
  y^{\alpha}(u_{0})}{du^{2}} }{|z-y(u_{0})|^{6}} \right. \nonumber \\
 & & \hspace{5mm} + \left( - q + \frac{q'}{2} \right)  \frac{ (z_{\mu} -
  y_{\mu}(u_{0})) \frac{d^{2} 
 y_{\nu}(u_{0})}{du^{2}}  + (z_{\nu} -  y_{\nu}(u_{0})) \frac{d^{2}
 y_{\mu} (u_{0})}{du^{2}} }{ |z-y(u_{0})|^{4}}  \nonumber \\  
  & & \hspace{5mm} + (- 4q + q' )   \frac{ \frac{ y_{\mu}(u_{0})}{du}
 \frac{ y_{\nu}(u_{0})}{du} 
  ( z_{\alpha} - y_{\alpha}(u_{0})) \frac{d^{2}
  y^{\alpha}(u_{0})}{du^{2}}  }{|z-y(u_{0})|^{4}} \nonumber \\
  & & \hspace{5mm} + 2q  g_{\mu \nu} \left. 
  \frac{ (z_{\alpha} - y_{\alpha}(u_{0})) \frac{ d^{2}
  y^{\alpha}(u_{0})}{du^{2}}}{|z-y(u_{0})|^{4}}  \right]
  W[C ]  + {\cal O}(z-y(u_{0}))^{-2}  \nonumber \\
 &+&  \frac{1}{N} \textrm{Tr} P w_{L, u_{0}} \frac{i}{4 \pi
  |z-y(u_{0})|^{3} } \left[ - (z_{\mu} - y_{\mu}(u_{0})) F_{\nu
  \alpha}(y(u_{0})) \frac{d y^{\alpha} (u_{0})}{du} \right. \nonumber
  \\ 
 & & \hspace{5mm} -  (z_{\nu} - y_{\nu}(u_{0}))
   F_{\mu \alpha}(y(u_{0}))  \frac{d y^{\alpha} (u_{0})}{du} 
   \nonumber \\
 & & \hspace{5mm} +  g_{\mu \nu} (z^{\alpha} - y^{\alpha}(u_{0}))
   F_{\alpha \beta}(y(u_{0})) \frac{d y^{\beta}(u_{0})}{du} \nonumber 
   \\
 & & \hspace{5mm} +  \left. (z^{\alpha} - y^{\alpha}(u_{0}))
  \left( F_{\mu \alpha} (y(u_{0})) \frac{d y_{\nu}(u_{0})}{du} +
  F_{\nu \alpha} (y(u_{0})) \frac{d y_{\mu}(u_{0})}{du}  \right)
  \right] w_{u_{0}, 0}  + \cdots \nonumber \\
 &+&  \frac{1}{N} \textrm{Tr} P w_{L, u_{0}} \left[ \frac{\theta_{i}(u_{0})
    \phi^{i}( y(u_{0}) ) }{24 \pi |z-y(u_{0})|^{3} } \left( 2 -
  \frac{\alpha}{10} + \frac{\beta}{2} \right) \right.  \nonumber \\
  & & \hspace{10mm} \times \left(  g_{\mu \nu} -
    3  \frac{(z_{\mu} - y_{\mu}(u_{0})) (z_{\nu} - y_{\nu}(u_{0}))
    }{|z-y(u_{0}) |^{2}} - \frac{d y_{\mu}(u_{0})}{du}
  \frac{dy_{\nu}(u_{0}) }{du}  \right)  \nonumber \\ 
  & & +   \frac{\theta_{i} (u_{0}) \phi^{i}(y(u_{0}))}{24 \pi
  |z-y(u_{0})|^{3}}  
    \left[  \left( 1 +  \frac{\alpha}{10} - \frac{\beta}{2} +
       \frac{\gamma}{5} \right) \left( (z_{\mu} - y_{\mu}(u_{0}))
  \frac{d^{2} y_{\nu}(u_{0}) }{du^{2}} +  (z_{\nu} - y_{\nu}(u_{0}))
  \frac{d^{2} y_{\mu}(u_{0}) }{du^{2}}  \right) \right. \nonumber \\
  & & \hspace{5mm} +  \left( 1 -  \frac{\alpha}{5} + \beta -
  \frac{\gamma}{5} \right) g_{\mu \nu} (z_{\alpha} -
    y_{\alpha} (u_{0})) \frac{d y^{\alpha}(u_{0})}{du^{2}} \nonumber
  \\
  & & \hspace{5mm} + \left( -3 + \frac{3 \alpha}{5} - 3 \beta - \frac{3
  \gamma}{5} \right)
  \frac{ (z_{\mu} - y_{\mu}(u_{0})) (z_{\nu} 
    - y_{\nu} (u_{0})) (z_{\alpha} - y_{\alpha} (u_{0}) ) \frac{d
    y^{\alpha}(u_{0})}{du^{2}} }{|z-y(u_{0})|^{3}} \nonumber \\
  & & \hspace{5mm} + \left. ( -3 + \gamma)  \frac{d
    y_{\mu}(u_{0}) }{du} \frac{d y_{\nu}(u_{0})}{du} (z_{\alpha} -
    y_{\alpha} (u_{0})) \frac{d^{2} y^{\alpha}(u_{0})}{du^{2}}  \right]
    \nonumber \\
  & & + \frac{ \frac{d \theta_{i}(u_{0}) }{du} }
    { 24 |z-y(u_{0})|^{3}}  \phi^{i}(y(u_{0})) \left( 2 + \frac{\alpha}{5}
  - \beta \right)  \left( (z_{\mu} -
    y_{\mu}(u_{0})) \frac{d 
    y_{\nu} (u_{0})}{du} + (z_{\nu} - y_{\nu} (u_{0})) \frac{d
    y_{\mu}(u_{0})}{ du}  \right) \nonumber \\
  & & + \frac{ \theta_{i}(u_{0})}{24 \pi |z-y(u_{0})|^{3}} 
   \left[ \left( -4 + \frac{\alpha}{5} \right) ( (z_{\mu} -
  y_{\mu}(u_{0})) D_{\nu} \phi^{i} ( y(u_{0}) ) + 
    (z_{\nu} - y_{\nu}(u_{0})) D_{\mu} \phi^{i}( y(u_{0}) ) )
  \right. \nonumber \\
 & & \hspace{5mm} + \left( 4 - \frac{\alpha}{5} \right)  g_{\mu \nu}
  (z_{\alpha} -  y_{\alpha}(u_{0}) ) D^{\alpha} \phi^{i} ( y(u_{0}) )
  \nonumber \\ 
  & & \hspace{5mm}  + \left( -6 - \frac{3 \alpha}{5} \right) 
  \frac{ (z_{\mu} -  y_{\mu}(u_{0})) (z_{\nu} -  
    y_{\nu}(u_{0})) (z_{\alpha} - y_{\alpha}(u_{0})) D^{\alpha}
    \phi^{i}( y(u_{0}) ) }{|z-y(u_{0})|^{2}} \nonumber \\
  & & \hspace{5mm} + (-2 + \alpha)   \frac{d
    y_{\mu}(u_{0}) }{du} \frac{d y_{\nu}(u_{0})}{du} (z_{\alpha} -
    y_{\alpha} (u_{0})) D^{\alpha} \phi^{i}(y(u_{0})) \nonumber \\
 & & \hspace{5mm} + \beta \frac{dy_{\alpha}(u_{0})}{du} (D^{\alpha} 
    \phi^{i}( y(u_{0}) ) ) \left( (z_{\mu} - y_{\mu}(u_{0})) \frac{d
    y_{\nu} (u_{0})}{du} \right. \nonumber \\
 & & \hspace{10mm}  + \left. \left.   (z_{\nu} - y_{\nu}
  (u_{0})) \frac{d  y_{\mu}(u_{0})}{ du}  \right)  \right]  \Biggr]
  w_{u_{0}, 0} +  \cdots. \label{AZ21final}
   \end{eqnarray}
 These free parameters may  depend on the coupling
 constant, and their meaning  is discussed in $\S$ 4. We have reached   
 the conclusion that the conformal Ward identity represents only the
 deformation of the Wilson loop and that the Wilson loop does not
 possess an 'anomalous dimension', irrespective of the values of the
  free parameters: 
  \begin{eqnarray}
 \textrm{(Translation)} & & \hspace{-3mm}  \int_{\cal M} d^{4}
   \partial^{\mu} z [ 
   T_{\mu \nu}(z) W[C ]   \xi^{\nu} ] = - \int ds \left( \frac{\delta
   W[C ]}{\delta y^{\nu}(s)} \right) \xi^{\nu}, \label{AZ21resboost} \\ 
 \textrm{(Dilatation)} & & \hspace{-3mm} \int_{\cal M} d^{4} z
   \partial^{\mu} [ T_{\mu 
   \nu} (z) W[C ]  \lambda z^{\nu} ] = - \int ds \left( \frac{\delta
   W[C ]}{\delta y^{\nu}(s)} \right) \lambda y^{\nu}(s),
   \nonumber \\
   \label{AZ21resdil} \\ 
  \textrm{(SCT)} & & \hspace{-3mm} \int_{\cal M} d^{4} z
   \partial^{\mu} [ T_{\mu \nu} (z) W[C ]  ( 2z^{\nu} (b_{\alpha}
   z^{\alpha}) - b^{\nu} z^{2} ) ] \nonumber \\ 
   &=& - \int ds \left( \frac{\delta W[C ]}{\delta y^{\nu}(s)} \right)
   (2 y^{\nu}(s) (b_{\alpha} y^{\alpha}(s)) - b^{\nu} (y(s))^{2}
   ). \label{AZ21ressct}
  \end{eqnarray}

 \section{OPE in the $U(1)$ SYM theory}
 In this section, we consider the OPE $ T_{\mu \nu}(z) W[C ] $ in
the $U(1)$ SYM theory in order to confirm the result of the previous
section by an explicit computation.  The bosonic part of the
Lagrangian of the $U(1)$ SYM theory is  
  \begin{eqnarray}
   {\cal L} &=&  \frac{1}{2 G^{2}} \left[ \frac{1}{2} \sqrt{g} g^{\mu
  \nu}  g^{\alpha \beta} F_{\mu \alpha}(z) F_{\nu \beta}(z)
  \right. \nonumber \\ 
  & & \hspace{5mm} +  \left.  \sqrt{g}
   g^{\mu \nu}  (\partial_{\mu} \phi_{i}(z)) (\partial_{\nu}
   \phi^{i}(z)) - \frac{1}{6} \sqrt{g} R \phi_{i}(z) \phi^{i}(z)
  \right],
   \label{AZM31action}  
  \end{eqnarray} 
 where the field strength is now $F_{\mu \nu}(z)  = \partial_{\mu}
 A_{\nu}(z) - \partial_{\nu} A_{\mu}(z)$. The terms proportional to
 the scalar curvature are necessary in order for the theory to be scale
 invariant. The energy-momentum tensor in the Euclidean flat space is
 \begin{eqnarray}
     T_{\mu \nu}(z) &=& \frac{1}{G^{2}} \left[ F_{\mu \alpha}(z)
   {F_{\nu}}^{\alpha}(z) - \frac{1}{4} g_{\mu \nu} F_{\alpha \beta}(z) 
   F^{\alpha \beta}(z)  \right] \nonumber \\
  &+& \frac{1}{G^{2}} \left[ (\partial_{\mu} \phi_{i}(z)) (\partial_{\nu}
   \phi^{i}(z)) - \frac{1}{2} g_{\mu \nu} (\partial_{\lambda}
   \phi_{i}(z)) (\partial^{\lambda} \phi^{i}(z)) \right] \nonumber \\
  &+&  \frac{1}{6G^{2}} [ -  \partial_{\mu} \partial_{\nu}
   (\phi_{i}(z) \phi^{i}(z)) +  g_{\mu  \nu} \Box (\phi_{i}(z)
   \phi^{i}(z))  ], \label{AZ31em} 
  \end{eqnarray}
 where $\Box$ denotes the Laplacian in flat space: $\Box =
 \partial^{\mu} \partial_{\mu}$.  We present a quick proof of
 (\ref{AZ31em}) in Appendix \ref{u1em}.  

 We adopt the Feynman gauge, in which the propagators are
  \begin{eqnarray}
    \langle A_{\mu}(z) A_{\nu}(w) \rangle  = \frac{G^{2}}{4 \pi^{2}}
         \frac{ g_{\mu \nu}}{(z-w)^{2}}, \hspace{3mm}
    \langle \phi_{i}(z) \phi_{j}(w) \rangle  =
       \frac{ G^{2}}{4 \pi^{2}} \frac{\delta_{ij}}{(z-w)^{2}}. \label{AZ3feyn}
  \end{eqnarray}
 
 The Wilson loop in the $U(1)$ SYM theory is
   \begin{eqnarray}
   W[C ] = \textrm{Tr} P \exp \left[ \oint_{C} du \left\{ i A_{\mu}(y(u))
   \frac{d y^{\mu} (u)}{du} + \theta_{i}(u) \phi^{i}(y(u)) \right\}
   \right],  \label{AZM2wilson} 
  \end{eqnarray} 
 where $u$ is the arc length parameter of the Wilson loop satisfying
 $|\frac{d y_{\mu}(u)}{du}|=1$ and $u$ is in the range  $0 \leq u \leq L$.
  $\theta_{i}(u)$ are again chosen to satisfy $\theta_{i}(u)
 \theta^{i}(u) = 1$.   

 We evaluate  the operator product $T_{\mu \nu}(z) W[C ]$  using
 Wick's theorem,  and then perform the integral by expanding the quantities
 about the $y_{\mu}(u_{0})$, the nearest point on the Wilson
 loop to $z_{\mu}$. Although the range of the arc length
 parameter of the Wilson loop is actually finite, we can approximate
 the Wilson loop by an infinitely long  straight line, because we
 are only concerned with  the situation in which the point $z_{\mu}$
 is in the vicinity of the Wilson loop in considering the conformal Ward
 identity. We expand the operator product  to ${\cal
 O}(z-y(u_{0}))^{-2}$. However,  
 for the contribution of $W[C ]$ itself, we compute only to  ${\cal
 O}(z-y(u_{0}))^{-3}$, because in this case the terms with
 singularities ${\cal O}(z-y(u_{0}))^{-2}$ have been 
 found not to contribute to the conformal Ward identity in the
 previous section. We give the detailed computation in Appendix
 \ref{u1abe}, and we only quote the result:
   \begin{eqnarray}
 & & T_{\mu \nu}(z) W[C ]   = \frac{G^{2}}{24
  \pi^{2} |z-y(u_{0})|^{4}} \left[ g_{\mu \nu} 
  - 2 \frac{dy_{\mu}(u_{0})}{du} \frac{dy_{\nu} (u_{0})}{du}
  \right. \nonumber   \\
 & & \hspace{5mm} - 2 \frac{(z_{\mu} - y_{\mu}(u_{0})) (z_{\nu} -
  y_{\nu}(u_{0})) } 
  { |z-y(u_{0})|^{2}}  \nonumber \\
 & & \hspace{5mm} -  2 \frac{(z_{\mu} - y_{\mu}(u_{0}) )(z_{\nu} -
  y_{\nu}(u_{0})) (z_{\alpha} - y_{\alpha} (u_{0}))}{|z-y(u_{0})|^{2}}
  \frac{d^{2} y^{\alpha}(u_{0})}{du^{2}} \nonumber \\
  & & \hspace{5mm}  -  \left( (z_{\mu} -
  y_{\mu}(u_{0})) \frac{d^{2} y_{\nu}(u_{0})}{du^{2}} + (z_{\nu} -
  y_{\nu}(u_{0})) \frac{d^{2} y_{\mu}(u_{0})}{du^{2}} \right)  \nonumber \\
 & &  \hspace{5mm} - 4 \frac{d y_{\mu}(u_{0})}{du} \frac{d
  y_{\nu}(u_{0})}{du}  (z_{\alpha} - y_{\alpha}(u_{0}) ) 
  \frac{d^{2} y^{\alpha} (u_{0}) }{du^{2}} 
  \nonumber \\
 & & \hspace{5mm}
  + 2 \left.  g_{\mu \nu} (z_{\alpha} - y_{\alpha}(u_{0})) \frac{d^{2}
  y^{\alpha}(u_{0})}{du^{2}} \right] W[C ] + {\cal O}(z-y(u_{0}))^{-2}
  \label{AZ22c} \\
 & & +  \frac{i}{4 \pi |z-y(u_{0})|^{3} } \left[ - (z_{\mu} -
 y_{\mu}(u_{0}))  F_{\nu \alpha} (y(u_{0})) \frac{d
 y^{\alpha}(u_{0})}{du} \right. \nonumber \\
 & & \hspace{5mm} -  (z_{\nu} - y_{\nu}(u_{0})) F_{\mu \alpha}
  (y(u_{0})) \frac{d y^{\alpha}(u_{0})}{du}  \nonumber \\  
  & & \hspace{5mm} + g_{\mu \nu} (z^{\alpha} -
 y^{\alpha}(u_{0}) ) F_{\alpha \beta}(y(u_{0})) \frac{d
 y^{\beta}(u_{0})}{du} \nonumber \\
  & & \hspace{5mm}  +  (z^{\alpha} - y^{\alpha}(u_{0}) )
 \left. \left( F_{\mu \alpha} (y(u_{0})) \frac{d y_{\nu}(u_{0})}{du} +
 F_{\nu \alpha} (y(u_{0})) \frac{d y_{\mu}(u_{0})}{du} \right) \right]
 W[C ] + \cdots  \nonumber \\ \label{AZ22vecres} \\
  &+&  \frac{\theta_{i}(u_{0}) \phi^{i}(y(u_{0})) }{24 \pi
    |z-y(u_{0})|^{3} } \left[ 2 g_{\mu \nu} - 
    6  \frac{(z_{\mu} - y_{\mu}(u_{0})) (z_{\nu} - y_{\nu}(u_{0}))
    }{|z-y(u_{0}) |^{2}} 
   - 2 \frac{d y_{\mu}(u_{0})}{du} 
    \frac{dy_{\nu}(u_{0}) }{du}  \right] W[C ] \nonumber \\
 &+&   \frac{\theta_{i} (u_{0}) \phi^{i}(y(u_{0})) }{24 \pi
    |z-y(u_{0})|^{3}}  
    \left[ \left( (z_{\mu} - y_{\mu}(u_{0})) \frac{d^{2}
    y_{\nu}(u_{0}) }{du^{2}} +  (z_{\nu} - y_{\nu}(u_{0})) \frac{d^{2}
    y_{\mu}(u_{0}) }{du^{2}}  \right) \right. \nonumber \\
  & & \hspace{5mm} +  g_{\mu \nu} (z_{\alpha} -
    y_{\alpha} (u_{0})) \frac{d y^{\alpha}(u_{0})}{du^{2}}  -3
  \frac{ (z_{\mu} - y_{\mu}(u_{0})) (z_{\nu} 
    - y_{\nu} (u_{0})) (z_{\alpha} - y_{\alpha} (u_{0}) ) \frac{d^{2}
    y^{\alpha}(u_{0})}{du^{2}} }{|z-y(u_{0})|^{3}} \nonumber \\
  & & \hspace{5mm}  -3   \left. \frac{d
    y_{\mu}(u_{0}) }{du} \frac{d y_{\nu}(u_{0})}{du} (z_{\alpha} -
    y_{\alpha} (u_{0})) \frac{d^{2} y^{\alpha}(u_{0})}{du^{2}} \right]
 W[C ]  \nonumber \\
  &+&  \frac{ \frac{d \theta_{i}(u_{0}) }{du} }
    { 12 |z-y(u_{0})|^{3}}  \phi^{i}(y(u_{0}))   \left[ (z_{\mu} -
    y_{\mu}(u_{0})) \frac{d 
    y_{\nu} (u_{0})}{du} + (z_{\nu} - y_{\nu} (u_{0})) \frac{d
    y_{\mu}(u_{0})}{ du}  \right] W[C ]  \nonumber \\
  &+&  \frac{ \theta_{i}(u_{0})}{24 \pi |z-y(u_{0})|^{3}} 
   \left[ -4 ( (z_{\mu} - y_{\mu}(u_{0})) \partial_{\nu}
 \phi^{i} ( y(u_{0}) ) + 
    (z_{\nu} - y_{\nu}(u_{0})) \partial_{\mu} \phi^{i}(y (u_{0})) )
    \right. \nonumber \\ 
 & & \hspace{5mm} + 4  g_{\mu \nu} (z_{\alpha} -
 y_{\alpha}(u_{0}) ) \partial^{\alpha} \phi^{i} (y (u_{0}) ) \nonumber 
    \\
 & & \hspace{5mm}   -6 \frac{ (z_{\mu} - y_{\mu}(u_{0})) (z_{\nu} - 
    y_{\nu}(u_{0})) (z_{\alpha} - y_{\alpha}(u_{0})) \partial^{\alpha}
    \phi^{i}( y(u_{0}) ) }{|z-y(u_{0})|^{2}} \nonumber \\
  & & \hspace{5mm} -2  \left.  \frac{d
    y_{\mu}(u_{0}) }{du} \frac{d y_{\nu}(u_{0})}{du} (z_{\alpha} -
    y_{\alpha} (u_{0})) \partial^{\alpha} \phi^{i}(y(u_{0}))
     \right] W[C ] + \cdots. \label{AZ22scares}
  \end{eqnarray}
 This result agrees with the general form (\ref{AZ21final}) 
 with the free parameters chosen as
  \begin{eqnarray}
   q = G^{2}, \hspace{2mm} q' = 0, \hspace{2mm} \alpha = \beta =
   \gamma = 0. \label{AZ3para} 
  \end{eqnarray}
 
 \section{Concluding remarks}
 We have investigated the OPE $ T_{\mu \nu}(z) W[C ]
$ for the ${\cal N}=4$, 4-dimensional $U(N)$ SYM theory. We have
pursued the general form of the OPE using  dimensional analysis and
the conservation law of the 
energy-momentum tensor. The general form of the Wilson loop is
given by (\ref{AZ21final}) with five free parameters undetermined.
We have obtained the following two results with
respect to the translation, dilatation and the special conformal
transformation. 
  \begin{itemize}
   \item{The Wilson loop does not possess an 'anomalous
       dimension'. This is analogous to the term
       $\frac{h}{(z-w)^{2}} A(w)$ 
       in the OPE of the 2-dimensional CFT, with $h$ being zero. In
       our case, however, the terms with  $W[C ]$ itself emerge
       non-trivially in the OPE as 
       (\ref{AZ21cnumresult4}) and (\ref{AZ21cnumresult3}), and yet
       they do not contribute to  the conformal Ward identity.}
   \item{The Wilson loop undergoes only a deformation. This result  
       is reminiscent of the term $\frac{1}{z-w} \partial A(w)$ in the
       2-dimensional CFT, which represents the replacement of the
       position of the operator $A(w)$. }
  \end{itemize}
 We have derived the explicit computation of the $U(1)$ SYM
 theory. This result reproduces the form obtained with the general
 analysis. In this case, the remaining free parameters are
 given in (\ref{AZ3para}).  

 Finally, we mention two future problems. The first is to pursue the
 OPE $ T_{\mu \nu}(z) W[C ] $ in terms of  supergravity in
 AdS space. In this context, the expectation value 
 of the Wilson loop is computed by considering the string world-sheet
 terminating on the loop $C$.  It is an interesting problem to extend
 this idea to the two-point  function $ T_{\mu \nu}(z) W[C ]
 $. 
 \begin{figure}
   \epsfxsize = 10cm
   \epsfysize = 5cm 
   \centerline{\epsfbox{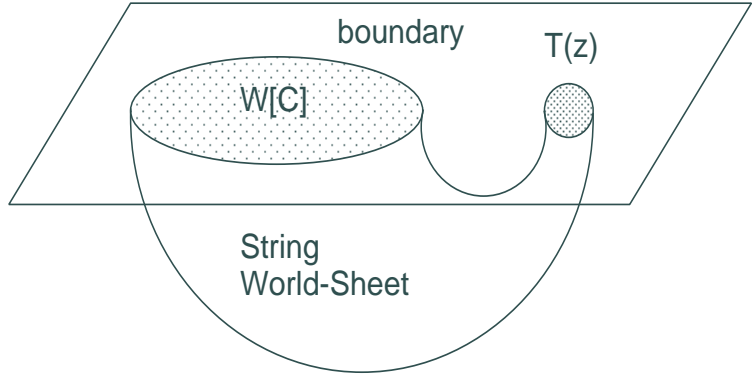}}
      \caption{We consider the two-point function  $ T_{\mu
        \nu}(z) W[C ] $ in the context of AdS/CFT.} 
   \label{supergr}
  \end{figure}

  The second problem is to apply our result to the computation of the
  expectation value of the Wilson loop, developed by Gross and
  Drukker. \cite{0010274} They have found that the expectation
  value of the Wilson loop stems from the conformal anomaly of the
  inversion by comparing the expectation value 
  of a straight line and that of a circular loop. We have attempted to
 interpret the anomaly of the special 
 conformal transformation, which maps a straight line into a circle, 
 in terms of the OPE $ T_{\mu \nu}(z) W[C ] $. The
 advantage of the OPE $ T_{\mu \nu}(z) W[C ] $ is that
 this is a gauge invariant quantity.  Although we have not yet obtained a
 definite answer, we conjecture that the undetermined coefficients $q$ 
  and $q'$ in the OPE (\ref{AZ21final}) may be related to the
  expectation value of 
 the Wilson loop. If we succeed in solidifying the interpretation of the
 conformal anomaly in terms of the OPE $ T_{\mu \nu}(z) W[C ]
 $, we may be able to gain insight into the gauge invariance of
 the result of Gross and Drukker. 

 \section*{Acknowledgments}
 We would like to thank Shin Nakamura for collaboration
 in the early stage of this project. We would also like to thank Hajime
 Aoki and Tsukasa Tada for useful discussions.

 \appendix 
  \section{Proof of (\ref{AZ23int1+}) $-$  (\ref{AZ23int3+})}
   \label{sp} 
   This appendix is devoted to the proof of the  frequently used
   formulae concerning the spherical integral.  

  \subsection{Integral in 3-dimensional Euclidean space}
  Before we consider the integral over the sphere in 4-dimensional
  space, we introduce the formulae concerning the
  integral over the surface of $S^{2}$ in 3-dimensional Euclidean
  space
   \begin{eqnarray}
     z_{1}^{2} + z_{2}^{2} + z_{3}^{2} = \epsilon^{2}.
   \end{eqnarray}
  The integrals over this sphere are given by 
  \begin{eqnarray}
   &(1')& \int_{S^{2}} d \Omega n_{a} = \int_{S^{2}} d \Omega n_{a}
   z_{b} z_{c} = 0, \label{AZMAint1} \\
   &(2')&  \int_{S^{2}} d \Omega n_{a} z_{b} = \frac{4 \pi
   \epsilon^{3}}{3} \delta_{ab}, \label{AZMAint2} \\
   &(3')& \int_{S^{2}} d \Omega n_{a} z_{b} z_{c} z_{d} = \frac{4 \pi
   \epsilon^{5}}{15} (\delta_{ab} \delta_{cd} + \delta_{ac}
   \delta_{bd} + \delta_{ad} \delta_{bc} ). \label{AZMAint3} 
   \end{eqnarray}

 \begin{wrapfigure}{r}{6.6cm}
   \epsfxsize = 5cm
   \epsfysize = 4.5cm 
   \centerline{\epsfbox{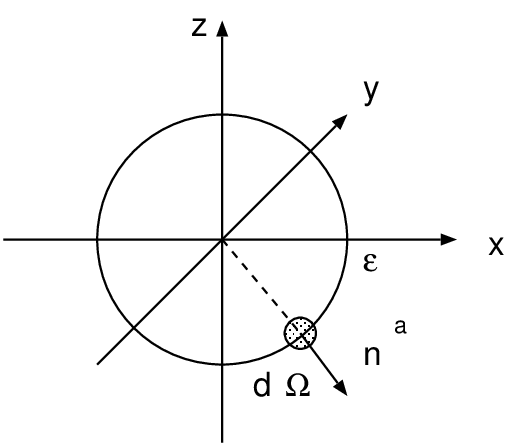}}
   \caption{We first compute the integral around the sphere as a
      simple example.}
   \label{spherepre}
   \end{wrapfigure}
 Here, the indices $a, b, c, \cdots$ run over $1$, $2$ and $3$, and
 $n^{a}$ denotes  
 the normal vector on the sphere. 
 These formulae are derived by noting the symmetry of the
  integral. 
  \begin{enumerate}
   \item{We first verify that the integrals (\ref{AZMAint1})
       are zero. The former 
       vanishes because of the symmetry of the sphere. For the
       latter integral, the result must be invariant under the $SO(3)$ 
       rotation of the integrand, and thus the result should be 
        \begin{eqnarray}
          \int_{S^{2}} d \Omega n_{a} z_{b} z_{c} \propto \epsilon_{abc}.
        \end{eqnarray}
       This manifestly vanishes because the integral is
       required to be symmetric under the exchange of $z_{b}$ and
       $z_{c}$. }
   \item{The key to the derivation of the integral (\ref{AZMAint2}) is
       to guess the final result from 
       the symmetry. This must give a non-zero result
       only when the indices satisfy $a=b$, and therefore the result
       is determined to be
        \begin{eqnarray}
          \int_{S^{2}} d \Omega n_{a} z_{b} = A \delta_{ab}
          \int_{S^{2}} d \Omega n_{c} z^{c}. 
        \end{eqnarray} 
      The coefficient $A$ is determined to be $\frac{1}{3}$ by
      contracting the indices $a$ and $b$. And the integral on the
      right-hand side is evaluated using Gauss's theorem:
       \begin{eqnarray}
        & & \int_{S^{2}} d \Omega n_{a} z_{b} = \frac{1}{3} \delta_{ab}
          \int_{S^{2}} d \Omega n^{c} z_{c} \nonumber \\
        &=& \frac{1}{3} \delta_{ab}
          \int_{B^{2}} d^{3} x \overbrace{\partial^{c}
          z_{c}}^{=1+1+1=3} = \frac{4 \pi \epsilon^{3}}{3} \delta_{ab},  
        \end{eqnarray} 
       where $B^{2}$ denotes the region inside the sphere $S^{2}$. }
   \item{The formula (\ref{AZMAint3}) is derived in much the same
       fashion. The 
       symmetry constrains the result to be invariant under the
       exchange of the indices $b, c$ and $d$, and this integral is
       determined to be
      \begin{eqnarray}
       \int_{S^{2}} d \Omega n_{a} z_{b} z_{c} z_{d} = A 
       ( \delta_{ab} 
       \delta_{cd} + \delta_{ac} \delta_{bd} + \delta_{ad} \delta_{bc} 
       ).
      \end{eqnarray}
     We are again left with the determination of the coefficient
     $A$. When we suppose that $a=b$ and that $c=d$, the both hand
     sides are rewritten as
    \begin{eqnarray}
    &\textrm{(l.h.s.)}& = \left( \int_{S^{2}} d \Omega n_{c}
     z^{c} \right) \epsilon^{2} = \left( \int_{B^{2}} d^{3} x
     \partial^{c} z_{c} \right)  
     \epsilon^{2} = 4 \pi \epsilon^{5}, \nonumber \\
    &\textrm{(r.h.s.)}& = A ( \delta_{aa} \delta_{cc} + 2 \delta_{ac}
     \delta_{ac} ) = A ( 3 \times 3 + 3 + 3) = 15A. \nonumber
  \end{eqnarray}
  Therefore, the coefficient $A$ is determined to be $\frac{4 \pi
  \epsilon^{5}}{15}$, and we obtain the result.}
  \end{enumerate}

 \subsection{ The integral over the enveloping surface}
  We next  extend the
  previous discussion to the integral over the enveloping surface
  wrapping the Wilson loop in the 4-dimensional Euclidean spacetime. 
 \begin{figure}
   \epsfxsize = 8cm
   \epsfysize = 5cm 
   \centerline{\epsfbox{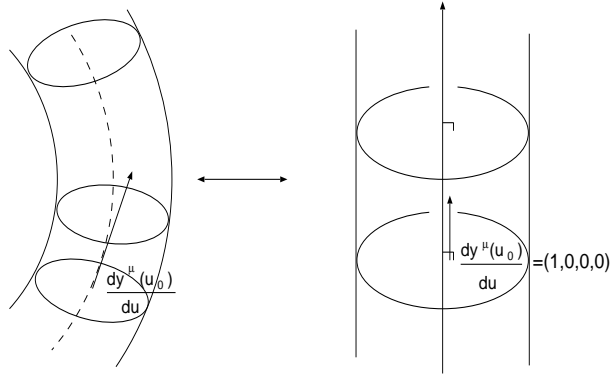}}
    \caption{We next evaluate the integral over the cylinder wrapping
      the Wilson loop.}
   \label{sphere}
 \end{figure}

  In the evaluation of the conformal Ward identity,  there frequently
  emerge the integrals (\ref{AZ23int1+}) $-$
  (\ref{AZ23int3+}).  These  are again derived by determining the
  forms through symmetry.  
  \begin{enumerate}
   \item{It is clear that the integrals (\ref{AZ23int1+}) vanish by
       noting the symmetry of the spherical integral.}
   \item{We verify (\ref{AZ23int2+}) through covariance. The result
       should be
      \begin{eqnarray}
       \int_{S^{2}(u_{0})} d \Omega n_{\mu} (z_{\nu} - y_{\nu}(u_{0}))
       = A g_{\mu \nu} + B \frac{d  y_{\mu}(u_{0})}{du} \frac{d
       y_{\nu}(u_{0})}{du}. 
      \end{eqnarray}
      The coefficients are determined by considering the special
      case. This problem is made simpler by considering the system 
      in which the Wilson line is a straight line parameterized by
        \begin{eqnarray} 
          y_{\mu} (u) = (u, 0, 0, 0),
        \end{eqnarray}
      as depicted in the right sketch in Fig. \ref{sphere}.
      The spherical integral thus reduces to the problem discussed in
      the previous section.
       \begin{itemize}
        \item{When $\mu = 0$, this integral vanishes, because the
            normal vector to the sphere is always perpendicular to the 
            tangent vector $\frac{d y_{\rho}(u_{0})}{du}$. This gives the
            constraint $A+B=0$ on the coefficients when $\nu$ is also
            0.} 
        \item{When neither $\mu$ nor $\nu$ is $0$, this reduces to the 
            integral (\ref{AZMAint2}), and the coefficient $A$ is
            determined as $A = \frac{4 \pi \epsilon^{3}}{3}$.} 
       \end{itemize}  
      Therefore, the coefficients are determined as $(A,B) = (\frac{4 \pi 
      \epsilon^{3}}{3}, - \frac{4 \pi \epsilon^{3}}{3})$, and the
      result (\ref{AZ23int2+}) is verified.}
   \item{The integral (\ref{AZ23int3+}) is verified in the same
       manner,  but this requires a somewhat tedious computation. It
       is clear that the result 
       must be invariant under the exchange of the indices $\nu, \rho, 
       \chi$. Also, the covariance constrains the result to be
       \begin{eqnarray}
   & & \hspace{-15mm} \int_{S^{2}(u_{0})} d \Omega n_{\mu} (z_{\nu} -
     y_{\nu}(u_{0})) 
     (z_{\rho} - y_{\rho}(u_{0})) (z_{\chi} - y_{\chi}(u_{0})) 
     \nonumber \\
     & & \hspace{-15mm} =   A  ( g_{\mu \nu} g_{\rho \chi} +  
     g_{\mu \rho} g_{\nu \chi} + g_{\mu \chi} g_{\nu \rho} )  +
     B \frac{d y_{\mu}(u_{0})}{du} \frac{d y_{\nu}(u_{0})}{du} \frac{d
     y_{\rho}(u_{0})}{du} \frac{d y_{\chi}(u_{0})}{du}
     \nonumber \\
    & & \hspace{-15mm} +  C  \left( g_{\mu \nu} \frac{d y_{\rho}(u_{0})}{du}
     \frac{d y_{\chi}(u_{0})}{du}  + g_{\mu \rho} \frac{d
     y_{\nu}(u_{0})}{du} \frac{d y_{\chi}(u_{0})}{du} + g_{\mu \chi}
     \frac{d y_{\nu}(u_{0})}{du} \frac{d y_{\rho}(u_{0})}{du} \right)
     \nonumber \\ 
    & & \hspace{-15mm} +D \left(  g_{\rho \chi}  \frac{d y_{\mu}(u_{0})}{du}
     \frac{d y_{\nu}(u_{0})}{du} +  g_{\nu \chi} \frac{d
     y_{\mu}(u_{0})}{du} \frac{d y_{\rho}(u_{0})}{du} +  g_{\nu \rho}
     \frac{d y_{\mu}(u_{0})}{du} \frac{d y_{\chi}(u_{0})}{du} \right).
     \nonumber \\
     \label{AZMAansatz3+}  
       \end{eqnarray}
     The coefficients $A,B,C$ and $D$ are again determined by considering
 the special case of the straight Wilson line. 
  \begin{itemize}
   \item{ When $\mu = 0$, the result again must be zero because the
       $S^{2}$ sphere is perpendicular to the $z_{0}$
       direction. Therefore, (\ref{AZMAansatz3+}) must vanish if we
       multiply it by $\frac{d y_{\mu}(u_{0})}{du} =
       \delta^{\mu}_{0}$: 
     \begin{eqnarray}
      & & (A+D) \left( \frac{d y_{\nu}(u_{0})}{du} g_{\rho 
       \chi} + \frac{d y_{\rho}(u_{0})}{du} g_{\chi 
       \nu}  + \frac{d y_{\chi}(u_{0})}{du}  g_{\nu 
       \rho} \right) \nonumber \\
    &+& (B + 3 C) \frac{d y_{\nu}(u_{0})}{du}\frac{d
       y_{\rho}(u_{0})}{du}  \frac{d y_{\chi}(u_{0})}{du} = 0.
     \end{eqnarray}
     This gives the constraints
     \begin{eqnarray}
      A + D = 0, \hspace{3mm} B + 3C = 0.
     \end{eqnarray}}
   \item{ When $\chi =0$ and $\mu, \nu, \rho = 1, 2, 3$, this must
       vanish, because of the relation (\ref{AZMAint1}). In this case,
       (\ref{AZMAansatz3+}) is rewritten, by multiplying
       $ \frac{d y_{\chi}(u_{0})}{du} =  \delta^{\chi}_{0}$, as 
     \begin{eqnarray}
  & & A \left( g_{\mu \nu}  \frac{d y_{\rho}(u_{0})}{du}
      +   g_{\nu \rho} \frac{d y_{\mu}(u_{0})}{du}
      +   g_{\rho \mu} \frac{d y_{\nu}(u_{0})}{du}
      \right)
  +   B   \frac{d y_{\mu}(u_{0})}{du} \frac{d y_{\nu}(u_{0})}{du}
      \frac{d y_{\rho}(u_{0})}{du}  \nonumber \\
  &+& C \left( g_{\mu \nu}  \frac{d y_{\rho}(u_{0})}{du} 
        + g_{\mu \rho} \frac{d y_{\nu}(u_{0})}{du}
        + \frac{d y_{\mu}(u_{0})}{du} \frac{d y_{\nu}(u_{0})}{du}
      \frac{d y_{\rho}(u_{0})}{du}  \right) \nonumber \\
  &+& D \left( 2 \frac{d y_{\mu}(u_{0})}{du} \frac{d y_{\nu}(u_{0})}{du}
      \frac{d y_{\rho}(u_{0})}{du}   + g_{\rho \nu} \frac{d
      y_{\mu}(u_{0})}{du}  \right) = 0. 
     \end{eqnarray}
   The coefficients are thus constrained to be 
    \begin{eqnarray}
    -A = C = D, \hspace{3mm} B + C + 2D = 0. 
    \end{eqnarray}}
  \item{Finally, we consider the case in which neither of the indices
      are $0$. In this case, $A$ is determined by identifying this
      case with the integral (\ref{AZMAint3}): $A = \frac{4 \pi
      \epsilon^{5}}{15}$. }
  \end{itemize}
  The coefficients in (\ref{AZMAansatz3+}) are thus determined to be
  \begin{eqnarray}
   (A, B, C, D) = \left( \frac{4 \pi \epsilon^{5}}{15}, \frac{4 \pi
   \epsilon^{5}}{5}, - \frac{4 \pi \epsilon^{5}}{15}, -\frac{4 \pi
   \epsilon^{5}}{15} \right).
  \end{eqnarray}  }
  \end{enumerate} 

\section{Proof of (\ref{AZ2wholedefloop}) } \label{defo}
 In this appendix, we present the explicit computation of the
 deformation of the Wilson loop (\ref{AZ2wholedefloop}). We deform the 
 loop slightly as
   \begin{eqnarray}
      y_{\mu}(s) \to y_{\mu}(s) + \delta y_{\mu}(s).
   \end{eqnarray}
 \begin{figure}
   \epsfxsize = 10cm
   \epsfysize = 1.7cm 
   \centerline{\epsfbox{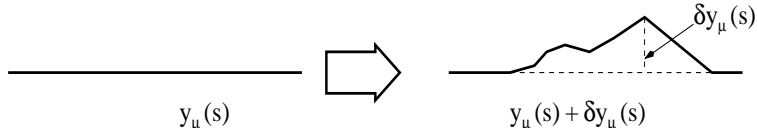}}
   \caption{The deformation of the loop.}
   \label{deformation}
 \end{figure}
  The  Wilson loop varies under this deformation as 
  \begin{eqnarray}
  W[C ] &\to& \frac{1}{N} \textrm{Tr} P \exp \left[ \int
  ^{2 \pi}_{0} ds \left\{ i 
  A_{\mu}(y(s) + \delta y(s)) \frac{d (y^{\mu} + \delta
  y^{\mu}(s))}{ds} \right. \right. \nonumber \\
  & & \hspace{5mm} + \left. \left. |\frac{d}{ds}(y_{\mu}(s) + \delta
  y_{\mu}(s))|  \theta_{i}(s) \phi^{i} (y(s) + \delta y(s)) \right\}
  \right] \nonumber \\ 
   &=& \frac{1}{N} \textrm{Tr} P \exp \left[ \int^{2 \pi}_{0} ds \left\{ i
  A_{\mu}(y(s))  \frac{dy^{\mu}(s)}{ds} + i (\partial_{\nu}
  A_{\mu}(y(s)) )  \frac{dy^{\mu}(s)}{ds} \delta y^{\nu}(s)
  \right. \right. \nonumber \\ 
  & & \hspace{5mm} + i A_{\nu}(y(s)) \frac{d (\delta
  y^{\nu}(s))}{ds} + \theta_{i}(s) \phi^{i}(y(s)) |\frac{d
  y_{\mu}(s)}{ds}| \nonumber \\
  & & \hspace{5mm} + \theta_{i}(s) (\partial_{\nu} \phi^{i}(y(s)))
  |\frac{d y_{\mu}(s)}{ds}| \delta y^{\nu}(s) \nonumber \\
  & & \hspace{5mm} + \left. \theta_{i}(s) \phi^{i}(y(s))
  \frac{1}{|\frac{dy_{\mu}(s)}{ds}|} 
  \frac{d y_{\nu}(s)}{ds} \left( \frac{d (\delta y^{\nu}(s))}{ds}
  \right)  \Bigr\} \right]  \nonumber \\ 
  &=& W[C ] + \int^{2 \pi}_{0} ds \delta y^{\nu}(s) \textrm{Tr} P
  {\hat w}_{2 \pi, s} \left[ i (\partial_{\nu} A_{\alpha}(y(s)) )
  \frac{d y^{\alpha}(s)}{ds} \right. \nonumber \\ 
  & & \hspace{5mm} + \left. 
  \theta_{i}(s) (\partial_{\nu} \phi^{i}(y(s)))
  |\frac{dy_{\mu}(s)}{ds}|  \right] {\hat w}_{s, 0} \nonumber \\
  & & - \frac{1}{N} \int^{2 \pi}_{0} ds \delta y^{\nu}(s) 
  \left[ \frac{d}{ds}  \left\{ \textrm{Tr} P ( {\hat  w}_{2 \pi, s} i
  A_{\nu}(y(s)){\hat w}_{s, 0} ) \right\}  \right] \nonumber \\
  & & - \frac{1}{N} \int^{2 \pi}_{0} ds \delta y^{\nu}(s) 
  \left[ \frac{d}{ds}  \left\{ \textrm{Tr} P \left(  {\hat  w}_{2 \pi, s}
  \theta_{i}(s) \phi^{i}(y(s)) 
  \frac{1}{|\frac{d y_{\mu}(s)}{ds}|} \frac{d y_{\nu}(s)}{ds}  {\hat
  w}_{s, 0} \right) \right\}  \right] \nonumber \\
  &=& W[C ] + \frac{1}{N} \int^{2 \pi}_{0} ds \delta y^{\nu}(s)
  \textrm{Tr} P {\hat
  w}_{2 \pi, s} \left[ i F_{\nu \alpha}(y(s)) \frac{d y^{\alpha}(s)}{ds}
  \nonumber + \theta_{i} 
  (s) |\frac{d y_{\mu}(s)}{ds}| D_{\nu} \phi^{i}(y(s))  \right] {\hat
  w}_{s, 0} \nonumber \\
  & & - \frac{1}{N} \int^{2 \pi}_{0} ds \delta y^{\nu}(s) 
  \left[ \frac{d}{ds}  \left\{ \textrm{Tr} P \left(  {\hat  w}_{2 \pi, s}
  \theta_{i}(s) \phi^{i}(y(s)) \frac{1}{|\frac{d y_{\mu}(s)}{ds}|}
  \frac{d y_{\nu}(s)}{ds}  {\hat  w}_{s, 0} \right) \right\} \right].
  \nonumber \\
  \end{eqnarray}
  We thus obtain the deformation of the Wilson loop as
 \begin{eqnarray}
  \left( \frac{\delta W[C ]}{\delta y^{\nu}(s)} \right) 
  &=&   \frac{1}{N} \textrm{Tr} P {\hat w}_{2 \pi, s} \left[  i F_{\nu
  \alpha} (y(s))  \frac{d y^{\alpha} (s)}{ds} + |\frac{dy_{\mu}(s)}{ds}|
  \theta_{i}(s)   D_{\nu} \phi^{i}( y(s) ) \right] {\hat w}_{s, 0}
  \nonumber \\ 
  &-& \frac{d}{ds} \left[ \frac{1}{N} \textrm{Tr} P {\hat w}_{2 \pi, s} 
  \left\{ \theta_{i}(s) \phi^{i}(y(s))
  \frac{ \frac{d y_{\nu}(s)}{ds}}{|\frac{d
  y_{\mu}(s)}{ds}| } \right\} {\hat w}_{s, 0} \right].  \nonumber
  \end{eqnarray}

\section{Proof of (\ref{AZ31em})} \label{u1em}
 This appendix provides a quick proof of the energy-momentum tensor
 of the $U(1)$ SYM theory, as an
 example of enjoying an advantage of the scale invariance and the
 conservation law of the energy-momentum tensor. The result 
 (\ref{AZ31em}) can be obtained by differentiating the scalar curvature
 with respect to the metric. Here, we take a shortcut by imposing  the
 following ansatz instead:
   \begin{eqnarray}
   & & \hspace{-5mm}  \frac{2}{\sqrt{g}} \frac{ - \frac{1}{12G^{2}}
    \sqrt{g} R \phi_{i}(z) 
    \phi^{i}(z)}{\partial g^{\mu \nu}}|_{\textrm{flat space}}
    \nonumber \\ 
   &=& \frac{1}{G^{2}} [ A \partial_{\mu} \partial_{\nu} (\phi_{i}(z)
    \phi^{i}(z)) + B g_{\mu \nu} \Box (\phi_{i}(z) \phi^{i}(z))  ].
   \end{eqnarray}
 The coefficients $A$ and $B$ should be determined by the
 tracelessness and divergencelessness condition of the energy-momentum
 tensor. The energy-momentum tensor in the Euclidean flat space is  
  \begin{eqnarray}
     T_{\mu \nu}(z) &=& \frac{2}{\sqrt{g}} \frac{\partial {\cal
   L}}{\partial g^{\mu \nu}}= \frac{1}{G^{2}} \left[ F_{\mu \alpha}(z)
   {F_{\nu}}^{\alpha}(z) - \frac{1}{4} g_{\mu \nu} F_{\alpha \beta}(z) 
   F^{\alpha \beta}(z) \right. \nonumber \\
 & & \hspace{-10mm} + (\partial_{\mu} \phi_{i}(z)) (\partial_{\nu}
   \phi^{i}(z)) - \frac{1}{2} g_{\mu \nu} (\partial_{\lambda}
   \phi_{i}(z)) (\partial^{\lambda} \phi^{i}(z)) \nonumber \\
 & & \hspace{-10mm} +  \left. A \partial_{\mu} \partial_{\nu}
   (\phi_{i}(z) \phi^{i}(z)) + B  g_{\mu  \nu} \Box (\phi_{i}(z)
   \phi^{i}(z))  \right]. 
  \end{eqnarray}
 These two fundamental properties of the energy momentum
 tensor impose the following constraints on the unknown coefficients. 
  \begin{itemize}
    \item{ ${T_{\mu}}^{\mu}(z) = 0$: This gives the relation
        $2A+8B-1=0$, with the help of the Klein-Gordon equation $\Box
        \phi_{i}(z) = 0$.}
   \item{ $\partial^{\mu} T_{\mu \nu}(z) = 0$: This leads to the
       constraint $A+B=0$.}
  \end{itemize}
  The coefficients are determined as $A = - \frac{1}{6}$ and
  $B = \frac{1}{6}$, and this completes the proof of (\ref{AZ31em}).

 \section{Proof of (\ref{AZ22c}) $-$ (\ref{AZ22scares})}
   \label{u1abe} 
 This appendix is devoted to the explicit computation of the OPE
 $ T_{\mu \nu}(z) W[C ] $ in the $U(1)$ SYM
 theory. We first take contractions following Wick's theorem.  Then,
 we obtain 
  \begin{eqnarray}
     T_{\mu \nu}(z) W[C ]  
  = ( T_{\mu \nu}(z) W[C ] )_{\textrm{c}}
  + ( T_{\mu \nu}(z) W[C ] )_{\textrm{vec}}
  + ( T_{\mu \nu}(z) W[C ] )_{\textrm{sca}}. \nonumber \\
  \end{eqnarray}
 Here, $( T_{\mu \nu}(z) W[C ] )_{\textrm{c}}$ represents the result of two
 contractions, while $( T_{\mu \nu}(z) W[C ] )_{\textrm{vec}}$ and $( T_{\mu
 \nu}(z) W[C ] )_{\textrm{sca}}$ represent the results of the single
 contraction with respect to the vector and scalar fields,
 respectively. 
 The explicit form of each component is given below. 
  \begin{eqnarray}
  ( T_{\mu \nu}(z) W[C ] )_{\textrm{c}} &=& G^{2}
    \oint_{C} du' du'' [ 
    -   ( \partial_{\alpha} D(z-y(u' )) )
        ( \partial^{\alpha} D(z-y(u'')) )
        {\dot y}_{\mu}(y') {\dot y}_{\nu}(u'') \nonumber \\
 & & \hspace{-20mm} -  ( \partial_{\mu} D(z-y(u' )) )
        ( \partial_{\nu} D(z-y(u'')) )
        {\dot y}_{\alpha}(u') {\dot y}^{\alpha}(u'') \nonumber \\ 
 & & \hspace{-20mm}   +   ( \partial_{\alpha} D(z-y(u' )) )
        ( \partial_{\nu}    D(z-y(u'')) )
        {\dot y}_{\mu}(u') {\dot y}^{\alpha}(u'') \nonumber \\
 & & \hspace{-20mm}
    +   ( \partial_{\mu}    D(z-y(u' )) )
        ( \partial^{\alpha} D(z-y(u'')) )
        {\dot y}_{\alpha}(u') {\dot y}_{\nu}(u'')  \nonumber \\
  & & \hspace{-20mm}   + \frac{g_{\mu \nu}}{2} (\partial^{\beta}
    D(z-y(u' )) ) (\partial_{\beta} D(z-y(u'')) )
      {\dot y}_{\alpha}(u') {\dot y}^{\alpha}(u'') \nonumber \\
  & & \hspace{-20mm}
   -  \frac{g_{\mu \nu}}{2} (\partial_{\alpha} D(z-y(u' )) )
      (\partial_{\beta}  D(z-y(u''))  )
      {\dot y}^{\alpha}(u'') {\dot y}^{\beta}(u')  ] W[C ] \nonumber \\
 & &  \hspace{-25mm} + G^{2} \oint_{C} du' du'' \theta_{k}(u')
    \theta^{k} (u'') \left[ \frac{1}{3}  (\partial_{\mu} 
   D(z-y(u' )) )( \partial_{\nu} D(z-y(u'')) ) \right.   \nonumber \\
  & & \hspace{-20mm} + \frac{1}{3} (\partial_{\nu} D(z-y(u' )) )
    ( \partial_{\mu}  D(z-y(u'')) )  \nonumber \\
  & & \hspace{-20mm}  - \frac{g_{\mu \nu}}{6} (\partial_{\alpha}
    D(z-y(u' )))  (\partial^{\alpha} D(z-y(u''))  ) \nonumber \\
 & & \hspace{-20mm} - \frac{1}{6}[ D(z-y(u'))  (\partial_{\mu}
   \partial_{\nu} D(z-y(u'')) ) + D(z-y(u''))  (\partial_{\mu}
   \partial_{\nu} D(z-y(u' )) )  ] \nonumber \\
 & & \hspace{-20mm} + \frac{1}{6} g_{\mu \nu}
  [ D(z-y(u')) (\delta(z-y(u'' ))) \nonumber \\
 & & \hspace{-15mm} + \left.  D(z-y(u'')) (\delta (z-y(u' )))  ] 
     \right] W[C ], \label{AZMBcnum} \\  
  ( T_{\mu \nu}(z) W[C ] )_{\textrm{vec}} &=&  i \left[ \oint_{C} du ( -
    (\partial_{\mu} D(z-y(u))) F_{\nu \alpha}(y(u)) 
   {\dot y}^{\alpha}(u) \right. \nonumber \\
 & & \hspace{-20mm} - (\partial_{\nu} D(z-y(u)))
     F_{\mu \alpha}(y(u)) {\dot y}^{\alpha}(u) \nonumber \\
 & & \hspace{-20mm} + g_{\mu \nu} (\partial^{\alpha} D(z-y(u)) )
   F_{\alpha \beta} (y(u))  {\dot y}^{\beta} (u) \nonumber \\
 & & \hspace{-20mm} \left.  + (\partial^{\alpha} D(z-y(u)) ) (F_{\mu 
    \alpha}(y(u)) {\dot y}_{\nu}(u) + F_{\nu \alpha}(y(u)) {\dot
    y}_{\mu}(u) ) )  \right] W[C ], \label{AZMBvec} \\
   ( T_{\mu \nu}(z) W[C ] )_{\textrm{sca}} &=& \oint_{C} du
   \frac{1}{3} \theta^{i}(u) \left[ \phi_{i}(y(u)) 
  (\partial_{\mu} \partial_{\nu} D(z-y(u)) ) \right. \nonumber \\
 & & \hspace{-20mm} - \left. \frac{1}{3} g_{\mu \nu} \phi_{i}(y(u))
    \delta (z-y(u)) \right] W[C ] \nonumber \\
 & & \hspace{-25mm}  + \oint_{C} du  \theta^{i}(u) \left[ 
  - \frac{2}{3}(\partial_{\mu} D(z-y(u)) ) (\partial_{\nu}
 \phi_{i}( y(u) ) )  - \frac{2}{3}(\partial_{\nu} D(z-y(u)) ) (\partial_{\mu}
   \phi_{i}( y(u) )) \right. \nonumber \\
  & & \hspace{-20mm} + \frac{g_{\mu \nu}}{3} ( \partial_{\alpha}
 D(z-y(u)) ) (\partial^{\alpha} \phi_{i}(y(u)) ) \nonumber \\
 & & \hspace{-20mm} + \frac{1}{3}(\partial^{\alpha} \phi_{i}(y(u)))
    (z_{\alpha} - y_{\alpha}(u)) 
 ( \partial_{\mu}  \partial_{\nu} D(z-y(u)) ) \nonumber \\
 & & \hspace{-20mm} \left.
 + \frac{1}{3} g_{\mu \nu} (\partial^{\alpha} \phi_{i}(y(u)) ) (z_{\alpha}
  - y_{\alpha}(u) ) \delta(z-y(u))  \right]  W[C ].
  \end{eqnarray}
In this appendix, the dotted quantities represents derivatives with
 respect to the arc length parameter: ${\dot y}_{\mu}(u) = \frac{d
 y_{\mu}(u)}{du}$, ${\ddot y}_{\mu}(u) = \frac{d^{2}
 y_{\mu}(u)}{du^{2}}$, and so on.
 Also, we define  $D(z-y(u)) = \frac{-1}{4 \pi^{2} (z-y(u))^{2}}$.

  We next perform the integral over the Wilson loop. We
  expand the quantities around the point $y_{\mu}(u_{0})$, the nearest
  point  on the Wilson loop to $z_{\mu}$. The
  miscellaneous quantities in the integral are expanded as follows:
  \begin{eqnarray}
   y_{\mu}(u) &=& y_{\mu}(u_{0}) + (u-u_{0}) {\dot y}_{\mu}(u_{0}) +
   \frac{(u-u_{0})^{2}}{2} {\ddot y}_{\mu} (u_{0}) + \cdots,
   \label{AZMBy}  \\
   \theta_{i}(u) &=& \theta_{i}(u_{0}) + (u-u_{0}) {\dot \theta}_{i}
   (u_{0}) + \frac{(u-u_{0})^{2}}{2} {\ddot \theta_{i}}(u_{0}) +
   \cdots, \label{AZMBtheta} \\
  \frac{1}{(z-y(u))^{2n}} &=&  \left[ (z_{\mu} - y_{\mu}(u_{0}) -
   (u-u_{0}) {\dot y}_{\mu} (u_{0}) - \frac{(u-u_{0})^{2}}{2} {\ddot
   y}_{\mu} (u_{0}) - \cdots )^{2} \right]^{-n}  \nonumber \\
  &=&  [ (z_{\mu} - y_{\mu}(u_{0}))^{2} +
   (u-u_{0})^{2}  ]^{-n}  \nonumber \\
  & & \times \left[ 1 - \frac{(u - u_{0})^{2} (z_{\mu} -
   y_{\mu}(u_{0})) {\ddot  y}^{\mu} (u_{0})}{ [ (z_{\mu} -
   y_{\mu}(u_{0}))^{2} +  (u-u_{0})^{2}  ]} - \cdots \right]^{-n}
   \nonumber \\ 
  &=& \frac{1}{[ (z_{\mu} - y_{\mu}(u_{0}))^{2} + (u-u_{0})^{2}  ]^{n}}
   \nonumber \\
  &+&  \frac{ n (u-u_{0})^{2}   (z_{\mu} -
   y_{\mu}(u_{0})) {\ddot  y}^{\mu} (u_{0}) }{[ (z_{\mu} -
   y_{\mu}(u_{0}))^{2} + (u-u_{0})^{2}  ]^{n+1} } + \cdots.
   \label{AZMBdeno} 
  \end{eqnarray}
    The expansion of the denominator (\ref{AZMBdeno})  needs
    some explanation. 
  \begin{itemize}
   \item{We expand the denominator $\frac{1}{(z-y(u))^{2n}}$ around 
    \begin{eqnarray}
    \frac{1}{[ (z_{\mu} - y_{\mu}(u_{0}))^{2} + (u-u_{0})^{2}  ]^{n}},
         \nonumber 
    \end{eqnarray}
         because we  approximate the integration over the Wilson loop by
         the integral over the straight line. In analyzing the
         conformal Ward identity, the point $z_{\mu}$ is on the
         $S^{2}(u_{0})$ sphere whose radius is $\epsilon$, and the
         point $z_{\mu}$ is in the vicinity of the Wilson
         loop. Therefore, the Wilson loop is perceived as a straight
         line, just as we perceive the 
         earth as a flat plane because the earth is much bigger than
         we are.}
   \item{In this expansion, we must be careful about the fact that
       ${\dot y}_{\alpha}(u_{0}) {\ddot y}^{\alpha}(u_{0}) =
       \frac{1}{2} (\frac{d}{du} |{\dot y}_{\alpha}(u_{0})|^{2} ) =
       0$ and that the definition of the point
       $y_{\mu}(u_{0})$ immediately leads to $(z_{\mu} -
       y_{\mu}(u_{0})) {\dot y}^{\mu}(u_{0})= 0$.}
  \end{itemize}
  The following integral is useful in the computation given below: 
   \begin{eqnarray}
    & & \hspace{-10mm} \int^{+\infty}_{-\infty} du
    \frac{(u-u_{0})^{n}}{2 \pi^{2} ((u-u_{0})^{2} + (z_{\mu} -
    y_{\mu}(u_{0}))^{2} )^{m} } \nonumber \\ 
   &=&  \{ \begin{array}{ll} 0 & (n \textrm{ is an odd number.}) \\ 
    \frac{ (n-1)!! (2m-n-3)!!}{2 \pi |z_{\mu} -
    y_{\mu}(u_{0}) |^{2m-n-1} (2m-2)!! } &  (n \textrm{ is an even
    number.}) \end{array},  \label{AZMBformula}
   \end{eqnarray}
 where $m$ and $n$ are positive integers satisfying $2m > n$. We
 now give the computation for deriving the formula (\ref{AZ22c}) $-$
 (\ref{AZ22scares}). The computation is rather lengthy,
 and we compute the terms one by one. 

 \subsection{ The computation of $( T_{\mu \nu}(z) W[C ]
 )_{\textrm{c}}$ } 
 Since this is a contribution of $W[C ]$ itself, it suffices to perform the
 computation to ${\cal O}(z-y(u_{0}))^{-3}$. We have already 
 shown that the terms of ${\cal O}(z-y(u_{0}))^{-2}$ do not contribute
 to the conformal Ward identity. The following
 quantities are frequently involved in  the computation: 
   \begin{eqnarray}
  & &  \int^{+\infty}_{-\infty} du {\dot y}_{\nu} (u) (\partial_{\mu}
    D(z-y(u) ) ) = O_{\mu \nu}^{-2} + O_{\mu \nu}^{-1} +
    \cdots, \nonumber  \\
  & & \int^{+\infty}_{-\infty} du \theta_{i}(u) (\partial_{\mu}
    D(z-y(u)) ) = P_{\mu}^{-2} + P^{-1}_{\mu} + \cdots, \nonumber \\ 
  & & \int^{+\infty}_{-\infty} du \theta_{i}(u) D(z-y(u)) =
    P^{-1} + P^{0} + \cdots, \nonumber \\
  & & \int^{+\infty}_{-\infty} du \theta_{i}(u) (\partial_{\mu}
    \partial_{\nu} D(z-y(u)) ) = P^{-3}_{\mu \nu} + P^{-2}_{\mu \nu} + 
    \cdots, \textrm{ where} \nonumber
  \end{eqnarray}
 \begin{eqnarray}
  O^{-2}_{\mu \nu} &=& \frac{(z_{\mu} - y_{\mu}(u_{0})) {\dot
  y}_{\nu}(u_{0})}{4 \pi |z-y(u_{0})|^{3}},  \\
  O^{-1}_{\mu \nu} &=& \frac{ [(z_{\alpha} - y_{\alpha}(u_{0})) {\ddot 
  y}^{\alpha}(u_{0}) ] (z_{\mu} - y_{\mu}(u_{0})) {\dot
  y}_{\nu}(u_{0})}{8 \pi |z-y(u_{0})|^{3}} \nonumber \\
   & & - \frac{{\dot
  y}_{\mu}(u_{0}) {\ddot y}_{\nu}(u_{0}) + \frac{1}{2} {\ddot
  y}_{\mu}(u_{0}) {\dot y}_{\nu} (u_{0})}{4 \pi |z-y(u_{0})|}, \\
  P_{\mu}^{-2} &=& \frac{ ( z_{\mu} - y_{\mu}(u_{0}) ) \theta_{i}
  (u_{0}) }{4 \pi |z-y(u_{0})|^{3} },  \\
  P_{\mu}^{-1} &=&  \frac{ [ (z_{\alpha} - y_{\alpha}(u_{0}) ) \cdot {\ddot
  y}^{\alpha}(u_{0})  ]  (z_{\mu} - y_{\mu}(u_{0}) ) \theta_{i}(u_{0})
  }{ 8 \pi |z-y(u_{0})|^{3} } \nonumber \\ 
  & & -  \frac{{\ddot y}_{\mu} (u_{0}) \theta_{i} (u_{0}) }{ 8 \pi
  |z-y(u_{0})| } 
   - \frac{ {\dot y}_{i} (u_{0}) {\dot \theta}_{i} (u_{0}) }{ 4 \pi
  |z-y(u_{0}) |  },  \\
  P^{-1} &=&  - \frac{ \theta_{i} (u_{0})}{ 4 \pi |z-y(u_{0})| },
  \nonumber \\ 
  P^{0} &=&  - \frac{ \theta_{i} (u_{0}) [ (z_{\alpha} -
  y_{\alpha}(u_{0})) {\ddot y}^{\alpha} (u_{0})  ]}{ 8 \pi |z-y(u_{0})|
  },  \\ 
  P^{-3}_{\mu \nu} &=& -  \frac{3 (z_{\mu} - y_{\mu}(u_{0}) ) (z_{\nu}
  - y_{\nu} (u_{0}) ) \theta_{i} (u_{0}) }{4 \pi |z-y(u_{0})|^{5} }
  \nonumber \\ 
  & & - \frac{ {\dot y}_{\mu}(u_{0}) {\dot y}_{\nu}(u_{0})
  \theta_{i}(u_{0}) }{ 4 \pi |z-y(u_{0})|^{3} } 
  + g_{\mu \nu} \frac{ \theta_{i}(u_{0})}{4 \pi |z-y(u_{0})|^{3}},
   \\ 
  P^{-2}_{\mu \nu} &=& - \frac{3 \theta_{i}(u_{0}) (z_{\mu} -
  y_{\mu}(u_{0}) ) (z_{\nu} - y_{\nu} (u_{0}) ) [ (z_{\alpha} - y_{\alpha}
  (u_{0}) ) {\ddot y}^{\alpha} (u_{0})  ] }{8 \pi  |z-y(u_{0})|^{5} }
  \nonumber \\ 
  & & + \frac{ {\dot \theta}_{i}(u_{0}) [ (z_{\mu} - y_{\mu} (u_{0}) )
  {\dot y}_{\nu} (u_{0}) + (z_{\nu} - y_{\nu} (u_{0})) {\dot y}_{\mu}
  (u_{0})  ]}{ 4 \pi |z-y(u_{0})|^{3} } \nonumber
  \\ 
  & & - \frac{ 3 \theta_{i} (u_{0}) [(z_{\alpha} - y_{\alpha} (u_{0}))
  {\ddot y}^{\alpha} (u_{0})  ] {\dot y}_{\mu} (u_{0}) {\dot y}_{\nu}
  (u_{0}) }{ 8 \pi |z-y(u_{0})|^{3}  } \nonumber \\
  & & + \frac{ \theta_{i} (u_{0}) [ (z_{\mu} - y_{\mu}(u_{0}) ) {\ddot
  y}_{\nu} (u_{0}) + (z_{\nu} - y_{\nu}(u_{0}) ) {\ddot y}_{\nu}
  (u_{0})  ]}{ 8 \pi |z-y(u_{0})|^{3} } \nonumber
  \\
  & & +   g_{\mu \nu} \frac{ \theta_{i}(u_{0}) [ (z_{\alpha} -
  y_{\alpha}(u_{0})) {\ddot y}^{\alpha}(u_{0}) ]}{ 8 \pi
  |z-y(u_{0})|^{3}  }.
 \end{eqnarray}
 Each term of (\ref{AZMBcnum}) is thus integrated as  follows: 
  \begin{eqnarray}
   & & \hspace{-10mm} - G^{2} \left[ \oint_{C} du' du''  
       ( \partial_{\alpha} D(z-y(u' )) )
        ( \partial^{\alpha} D(z-y(u'')) )
        {\dot y}_{\mu}(u') {\dot y}_{\nu}(u'')  \right] W[C ] \nonumber \\
  &=&  - G^{2} ( {O^{-2 \alpha}}_{\mu} + {O^{-1 \alpha}}_{\mu} + \cdots )
       (O^{-2}_{\alpha \nu} + O^{-1}_{\alpha \nu} + \cdots  ) W[C ]
       \nonumber \\
  &=& -  G^{2} ( {O^{-2 \alpha}}_{\mu} O^{-2}_{\alpha \nu} ) W[C ] -  
  G^{2}  ( {O^{-2 \alpha}}_{\mu} O^{-1}_{\alpha \nu} + {O^{-2
       \alpha}}_{\nu}  O^{-1}_{\alpha \mu} ) W[C ] + \cdots \nonumber \\
  &=&  - \frac{G^{2}}{16 \pi^{2}} \frac{ {\dot y}_{\mu}
   (u_{0}) {\dot y}_{\nu}(u_{0})}{ |z-y(u_{0})|^{4}} W[C ] 
      + {\cal O}(z-y(u_{0}))^{-2}, \\
  & & \hspace{-10mm} - G^{2} \left[ \oint_{C} du' du'' ( \partial_{\mu}
       D(z-y(u' )) ) 
        ( \partial_{\nu} D(z-y(u'')) )
        {\dot y}_{\alpha}(u') {\dot y}^{\alpha}(u'')  \right] W[C ]
       \nonumber \\ 
  &=& -  G^{2} ( {{O^{-2}}_{\mu}}^{\alpha} O^{-2}_{\nu \alpha} ) W[C ] - 
    G^{2} ( {{O^{-2}}_{\mu}}^{\alpha} O^{-1}_{\nu \alpha} +
       {{O^{-2}}_{\nu}}^{\alpha}  O^{-1}_{\mu \alpha} ) W[C ] + \cdots
       \nonumber \\ 
  &=& \left[  - \frac{G^{2}}{16 \pi^{2}} \frac{(z_{\mu} -
  y_{\mu}(u_{0})) (z_{\nu} - y_{\nu}(u_{0}))}{|z-y(u_{0})|^{6}}
     \right.   \nonumber \\
  & &  - \frac{G^{2}}{16 \pi^{2}} \frac{ (z_{\alpha} -
  y_{\alpha}(u_{0})){\ddot y}^{\alpha} (u_{0}) (z_{\mu} -
  y_{\mu}(u_{0}) )(z_{\nu} - y_{\nu}(u_{0}) )}{ |z-y(u_{0})|^{6} }
       \nonumber \\
  & & + \left. \frac{G^{2}}{ 32 \pi^{2}} \frac{ (z_{\mu} - y_{\mu}(u_{0}))
  {\ddot y}_{\nu}(u_{0}) +  (z_{\nu} - y_{\nu}(u_{0})) {\ddot y}_{\mu}
  (u_{0}) }{|z-y(u_{0})|^{4}  }  \right] W[C ] \nonumber \\
  & & + {\cal O}(z-y(u_{0}))^{-2}, \\ 
  & & \hspace{-10mm}  G^{2} \left[  \oint_{C} du' du''
    [ ( \partial_{\alpha} D(z-y(u' )) )
        ( \partial_{\nu}    D(z-y(u'')) )
        {\dot y}_{\mu}(u') {\dot y}^{\alpha}(u'') \right.\nonumber \\ 
  & & \hspace{5mm} + \left. ( \partial_{\mu}    D(z-y(u' )) )
        ( \partial^{\alpha} D(z-y(u'')) )
        {\dot y}_{\alpha}(u') {\dot y}_{\nu}(u'') ] \right] W[C ]
       \nonumber \\ 
  &=& [ G^{2} ( {O^{-2 \alpha}}_{\mu} O^{-2}_{\nu \alpha} )  +
   G^{2} ( {O^{-2 \alpha}}_{\mu} O^{-1}_{\nu \alpha} +
       {{O^{-2}}_{\nu}}^{\alpha}  O^{-1}_{\alpha \mu} ) \nonumber \\
  & & \hspace{5mm}
   +  G^{2} ( {O^{-2 \alpha}}_{\nu} O^{-2}_{\mu \alpha} ) +
    G^{2} ( {O^{-2 \alpha}}_{\nu} O^{-1}_{\mu \alpha} +
       {{O^{-2}}_{\mu}}^{\alpha} O^{-1}_{\alpha \nu} )  ] W[C ] + \cdots
       \nonumber \\ 
  &=& \left[ - \frac{G^{2}}{8 \pi^{2}} \frac{ ((z_{\alpha} -
  y_{\alpha}(u_{0})) {\ddot y}^{\alpha} (u_{0}) ) {\dot
  y}_{\mu}(u_{0}) {\dot y}_{\nu}(u_{0})  }{ |z-y(u_{0})|^{4} }
   \right.    \nonumber \\
  & & - \left. \frac{G^{2}}{16 \pi^{2}} \frac{ (z_{\mu} -
  y_{\mu}(u_{0}) ) {\ddot y}_{\nu}(u_{0}) + (z_{\nu} - y_{\nu}(u_{0})) 
  {\ddot y}_{\mu}(u_{0}) }{|z-y(u_{0})|^{4} } \right] W[C ]  \nonumber
       \\ 
  & &  + {\cal O}(z-y(u_{0}))^{-2}, \\
  & & \hspace{-10mm}  \frac{G^{2}}{2} \left[ \oint_{C} du' du''
       g_{\mu \nu} (\partial^{\beta} D(z-y(u' )) ) 
      (\partial_{\beta} D(z-y(u'')) )
      {\dot y}_{\alpha}(u') {\dot y}^{\alpha}(u'') \right] W[C ]
       \nonumber \\  
  &=& \frac{G^{2}}{2} g_{\mu \nu} [  (O^{-2}_{\alpha \beta}
  O^{-2 \alpha  \beta} ) + 2 (O^{-2}_{\alpha \beta} O^{-1 \alpha
  \beta} )  ] W[C ] + \cdots \nonumber \\
  &=&  \frac{G^{2}}{ 32 \pi^{2}} \frac{g_{\mu
  \nu}}{|z-y(u_{0})|^{4}} W[C ] + {\cal O}(z-y(u_{0}))^{-2},  \\ 
  & & \hspace{-10mm}  - \frac{G^{2}}{2} \left[ \oint_{C} du' du''
       g_{\mu \nu} (\partial_{\alpha} D(z-y(u' )) ) 
      (\partial_{\beta}  D(z-y(u''))  )
      {\dot y}^{\alpha}(u'') {\dot y}^{\beta}(u')  \right] W[C ]
       \nonumber \\ 
  &=&  - \frac{G^{2}}{2} g_{\mu \nu} [  (O^{-2}_{\alpha \beta}
  O^{-2 \beta  \alpha} ) + 2 (O^{-2}_{\alpha \beta} O^{-1 \beta
  \alpha} )  ] W[C ] + \cdots \nonumber \\ 
  &=&  \frac{G^{2}}{16 \pi^{2}} g_{\mu \nu} \frac{ (z_{\alpha} -
  y_{\alpha}(u_{0})) {\ddot y}^{\alpha}(u_{0}) }{|z-y(u_{0})|^{4} }
       W[C ]  + {\cal O}(z-y(u_{0}))^{-2},  \\
  & & \hspace{-10mm}   \frac{G^{2}}{3} \left[ \oint_{C} du' du''
       \theta_{i}(u') 
    \theta^{i} (u'')  [ (\partial_{\mu} 
   D(z-y(u' )) )( \partial_{\nu} D(z-y(u'')) )  \right. \nonumber \\
  & & \hspace{5mm}  + \left. (\partial_{\nu} D(z-y(u' )) )( \partial_{\mu}
      D(z-y(u'')) )  ]  \right] W[C ] \nonumber \\
  &=& \frac{2 G^{2}}{3}
  ( P^{-2}_{\mu} + P^{-1}_{\mu} + \cdots )( P^{-2}_{\nu}   +
       P^{-1}_{\nu} + \cdots ) W[C ] \nonumber \\ 
  &=&  \frac{ 2 G^{2}}{3}
  [ ( P^{-2}_{\mu} P^{-2}_{\nu} ) + (P^{-2}_{\mu} P^{-1}_{\nu} +
  P^{-2}_{\nu} P^{-1}_{\mu} ) + \cdots  ] W[C ] \nonumber \\
  &=&  \left[ \frac{ G^{2}}{24 \pi^{2}} \frac{ (z_{\mu} -
  y_{\mu}(u_{0})) (z_{\nu} - y_{\nu} (u_{0}) ) }{ |z_{\mu} - y_{\mu}
  (u_{0}) |^{6}}  \right. \nonumber \\
  & &  + \frac{G^{2}}{24 \pi^{2}} \frac{ [( z_{\alpha} - y_{\alpha}
  (u_{0}) ){\ddot y}^{\alpha} (u_{0})  ] (z_{\mu} - y_{\mu} (u_{0}))
  (z_{\nu} - y_{\nu} (u_{0}) ) }{ |z-y(u_{0})|^{6} } \nonumber \\
  & & - \left. \frac{G^{2}}{48 \pi^{2}} \frac{ (z_{\mu} - y_{\mu}(u_{0})) 
  {\ddot y}_{\nu} (u_{0}) + (z_{\nu} - y_{\nu} (u_{0})) {\ddot
  y}_{\mu} (u_{0}) }{ |z-y(u_{0})|^{4}} \right] W[C ] \nonumber \\
  & & + {\cal O}(z-y(u_{0}))^{-2},
    \\ 
  & & \hspace{-10mm} -   \frac{G^{2}}{6} \left[ \oint_{C} du' du'' g_{\mu
       \nu}  \theta_{i}(u') \theta^{i} (u'') (\partial_{\alpha} 
    D(z-y(u' )))  (\partial^{\alpha} D(z-y(u''))  )  \right] W[C ]
       \nonumber \\ 
  &=&  - \frac{G^{2}}{6} g_{\mu \nu}
  ( P^{-2}_{\alpha} + P^{-1}_{\alpha} + \cdots )
  ( P^{-2 \alpha} + P^{-1 \alpha} + \cdots ) W[C ] \nonumber \\ 
  &=& - \frac{G^{2}}{6} g_{\mu \nu} [ (P^{-2}_{\alpha} P^{-2
  \alpha} ) + (2 P^{-2}_{\alpha} P^{-1 \alpha})  ] W[C ] + \cdots  
  \nonumber \\
 &=& - \frac{G^{2}}{96 \pi^{2}} g_{\mu \nu} \frac{1}{|z-y(u_{0})|^{4}} W[C ] 
  + {\cal O}(z-y)^{-2}, \\
 & & \hspace{-10mm} -   \frac{G^{2}}{6}  \left[ \oint_{C} du' du''
       \theta_{i}(u') \theta^{i} (u'') [ D(z-y(u'))  (\partial_{\mu}
   \partial_{\nu} D(z-y(u'')) ) \right. \nonumber \\
 & & \hspace{5mm} + \left. D(z-y(u''))  (\partial_{\mu}
   \partial_{\nu} D(z-y(u' )) )  ]  \right]  W[C ] \nonumber \\
 &=&  - \frac{G^{2}}{3} (P^{-1} + P^{0}  +\cdots ) (P^{-3}_{\mu  
  \nu} + P^{-2}_{\mu \nu} + \cdots ) W[C ] \nonumber \\
 &=& - \frac{G^{2}}{3}
  [ (P^{-1} P^{-3}_{\mu \nu} ) + (P^{0} P^{-3}_{\mu \nu} + P^{-1}
  P^{-2}_{\mu \nu} )  ] W[C ] + \cdots \nonumber \\
  &=& \left[ \frac{G^{2}}{48 \pi^{2} } \left[ -3  \frac{ (z_{\mu} -
       y_{\mu} (u_{0})) (z_{\nu} - y_{\nu} (u_{0})) }{ |z-y(u_{0})|^{6}} 
   - \frac{{\dot y}_{\mu}(u_{0}) {\dot y}_{\nu}(u_{0}) }
       { |z-y(u_{0})|^{4} }  
   + g_{\mu \nu} \frac{1}{|z-y(u_{0})|^{4}}  \right] \right. \nonumber \\
  & & - \frac{G^{2}}{16 \pi^{2}} \frac{(z_{\mu} - y_{\mu}(u_{0})) 
  (z_{\nu} - y_{\nu} (u_{0})) [ (z_{\alpha} - y_{\alpha}(u_{0}))
  {\ddot y}^{\alpha}(u_{0})  ] }{ |z-y(u_{0})|^{6} } \nonumber \\
  & &  -  \frac{G^{2}}{24 \pi^{2}} \frac{ [(z_{\alpha} - y_{\alpha} 
  (u_{0}) ) {\ddot y}^{\alpha} (u_{0}) ] {\dot y}_{\mu} (u_{0}) {\dot
  y}_{\nu} (u_{0}) }{ |z-y(u_{0})|^{4} } \nonumber \\
  & & + \frac{G^{2}}{96 \pi^{2}} \frac{ (z_{\mu} - y_{\mu}
  (u_{0})) {\ddot y}_{\nu}(u_{0}) + (z_{\nu} - y_{\nu}(u_{0})) {\ddot
  y}_{\mu} (u_{0}) }{ |z-y(u_{0})|^{4} } \nonumber \\
  & & + \left. \frac{G^{2}}{48 \pi^{2}} g_{\mu \nu} \frac{ (z_{\alpha} -
  y_{\alpha} (u_{0}) ) {\ddot y}^{\alpha}(u_{0}) }{ |z-y(u_{0})|^{4}}
   \right] W[C ]  + {\cal O}(z-y(u_{0}))^{-2}. 
  \end{eqnarray}
 Above, we have utilized the fact that the term $\theta_{i}(u_{0}) {\dot
 \theta}^{i}(u_{0})$ vanishes because we have set $\theta_{i}(u_{0})
 \theta^{i}(u_{0}) = 1$. Summing all of the these terms, we obtain the 
 result (\ref{AZ22c}). 

 \subsection{The computation of $( T_{\mu \nu}(z) W[C ]
 )_{\textrm{vec}}$} 
  This computation is trivial, because the singularity of the
  integrand is ${\cal  O}(z-y(u))^{-3}$. Only the leading term in the
  expansion around the nearest point survives, and we can trivially
  read off the result (\ref{AZ22vecres}). 

  \subsection{The computation of $( T_{\mu \nu}(z) W[C ]
 )_{\textrm{sca}}$} 
 We next compute the effect of the scalar field. The following three
 terms are computed  as easily as those of the vector field:
  \begin{eqnarray}
  & &   \oint_{C} du  \theta^{i}(u) \left[ 
  - \frac{2}{3}(\partial_{\mu} D(z-y(u)) ) (\partial_{\nu}
 \phi_{i}( y(u) ) )
  - \frac{2}{3}(\partial_{\nu} D(z-y(u)) ) (\partial_{\mu}
   \phi_{i}( y(u) )) \right. \nonumber \\
  & &  + \left. \frac{g_{\mu \nu}}{3} ( \partial_{\alpha}
 D(z-y(u)) ) (\partial^{\alpha} \phi_{i}(y(u)) )  \right]  W[C ]
  \nonumber \\ 
  &=& - \frac{1}{6 \pi |z-y(u_{0})|^{3} } \theta_{i}(u_{0})
   [ (z_{\mu} - y_{\mu} (u_{0}) ) (\partial_{\nu} \phi^{i} (y(u_{0}) 
   ) ) \nonumber \\
  & & \hspace{5mm} + (z_{\nu} - y_{\nu}(u_{0}) ) (\partial_{\mu} \phi^{i}
   (y(u_{0}) ) )  ] W[C ] \nonumber \\
   & & +\frac{1}{12 \pi |z-y(u_{0})|^{3} } g_{\mu \nu} \theta_{i} (u_{0}) 
   (z_{\alpha} - y_{\alpha} (u_{0}) ) (\partial^{\alpha} \phi^{i}
   ( y(u_{0}) ) ) W[C ]. \label{AZMBsca1}
  \end{eqnarray}

 However, there are two terms that require a non-trivial computation.
 First, we compute the term in which the singularity of the 
 integrand is ${\cal O}(z-y(u))^{-4}$:
  \begin{eqnarray}
 & &   \frac{G^{2}}{3} \left[ \oint_{C} du \theta^{i}(u)
  \phi_{i}(y(u))   (\partial_{\mu} \partial_{\nu} D(z-y(u)) ) \right]
  W[C ]  \nonumber \\ 
 &=&  \frac{G^{2}}{3} \int^{+\infty}_{-\infty} du \theta_{i}(u_{0})
  \phi^{i}(y (u_{0})) \left[ \frac{g_{\mu \nu}}{2
  \pi^{2} [ (z_{\mu} - y_{\mu}(u_{0}))^{2} + (u-u_{0})^{2} ]^{2} }
  \right. \nonumber \\
 & & \hspace{5mm} +
  \frac{2 g_{\mu \nu} (u-u_{0})^{2} (z_{\alpha} - y_{\alpha}(u_{0}))
  {\ddot y}^{\alpha} (u_{0}) }{2 \pi^{2} [ (z_{\mu} -
  y_{\mu}(u_{0}))^{2} + (u-u_{0})^{2} ]^{3} } \nonumber \\
  & & \hspace{5mm} - \frac{4 (z_{\mu} - y_{\mu}(u_{0})) (z_{\nu} -
  y_{\nu}(u_{0})) }{2 \pi^{2}[ (z_{\mu} - y_{\mu}(u_{0}))^{2} +
  (u-u_{0})^{2} ]^{3} } -  \frac{12 (u-u_{0})^{2} (z_{\mu} -
  y_{\mu}(u_{0})) (z_{\nu} - y_{\nu}(u_{0})) }{2 \pi^{2} [ (z_{\mu} -
  y_{\mu}(u_{0}))^{2} +  (u-u_{0})^{2} ]^{4} } \nonumber \\
  & & \hspace{5mm} - \frac{4 {\dot y}_{\mu}(u_{0}) {\dot
  y}_{\nu}(u_{0}) (u-u_{0})^{2}}{2 \pi^{2}  [ (z_{\mu} -
  y_{\mu}(u_{0}))^{2} + (u-u_{0})^{2} ]^{3} }  \nonumber \\
  & & \hspace{5mm} - \frac{12 (z_{\alpha} - y_{\alpha}(u_{0}))
  {\ddot y}^{\alpha} (u_{0}) {\dot y}_{\mu}(u_{0}) {\dot
  y}_{\nu}(u_{0}) (u-u_{0})^{2}}{2 \pi^{2}  [ (z_{\mu} -
  y_{\mu}(u_{0}))^{2} + (u-u_{0})^{2} ]^{4} } \nonumber \\
  & & \hspace{5mm} + \left.  \frac{2  (u-u_{0})^{2} [ (z_{\mu} -
  y_{\mu}(u_{0})) {\ddot y}_{\nu} (u_{0}) + (z_{\nu} - y_{\nu}(u_{0}))
  {\ddot  y}_{\mu} (u_{0})  ] }{2 \pi^{2}  [ (z_{\mu} -
  y_{\mu}(u_{0}))^{2} + (u-u_{0})^{2} ]^{3} } + \cdots  \right] W[C ]
  \nonumber \\  
  & & +   \frac{G^{2}}{3} \int^{+\infty}_{-\infty} du \left[ \frac{d}{du}
  ( \theta_{i}(u_{0}) \phi^{i}( y(u_{0})) )  \right] \nonumber \\
  & & \hspace{5mm} \times  \frac{4 (u-u_{0})^{2}
  [ (z_{\mu} - y_{\mu}(u_{0})) {\dot y}_{\nu}(u_{0}) +  (z_{\nu} -
  y_{\nu}(u_{0})) {\dot y}_{\mu}(u_{0})  ]}{ 2 \pi^{2}  [ (z_{\mu} -
  y_{\mu}(u_{0}))^{2} + (u-u_{0})^{2} ]^{3}} W[C ] \nonumber \\
  &=&  \theta_{i}(u_{0}) \phi^{i} (y(u_{0}))
   \left[ \frac{g_{\mu \nu} 
   }{12 \pi |z-y(u_{0})|^{3} } - \frac{(z_{\mu} - y_{\mu}(u_{0}))
   (z_{\nu} - y_{\nu}(u_{0}) ) }{ 4 \pi |z-y(u_{0})|^{5}}
  \right. \nonumber \\ 
  & & \hspace{5mm} - \left. \frac{
   {\dot y}_{\mu}(u_{0}) {\dot y}_{\nu}(u_{0}) }{ 12 \pi
   |z-y(u_{0})|^{3} |{\dot y}(u_{0})|^{2}} \right] W[C ] \nonumber \\
  & & +  \frac{1}{24 \pi} g_{\mu \nu} \theta_{i}(u_{0}) \phi^{i}
   (y(u_{0})) \frac{ (z_{\alpha} - y_{\alpha}(u_{0})) {\ddot
   y}^{\alpha} (u_{0}) }{ |z-y(u_{0})|^{3}  } W[C ]
   \nonumber \\
   & & -  \frac{1}{8 \pi} \theta_{i}(u_{0}) \phi^{i} (y(u_{0})) 
   \frac{ (z_{\mu} - y_{\mu}(u_{0}) ) (z_{\nu} - y_{\nu}(u_{0}) )
   [(z_{\alpha} - y_{\alpha}(u_{0})) {\ddot y}^{\alpha}(u_{0})  ]}
   { |z-y(u_{0})|^{5} } W[C ] \nonumber \\
 & & - \frac{1}{8 \pi} \theta_{i}(u_{0}) \phi^{i} (y(u_{0})) 
   \frac{ {\dot y}_{\mu}(u_{0}) {\dot y}_{\nu}(u_{0}) [(z_{\alpha} -
   y_{\alpha} (u_{0}) ) {\ddot y}^{\alpha}(u_{0})  ]}{ |z-y(u_{0})|^{3} 
    } W[C ] \nonumber \\
 & & + \frac{1}{24 \pi} \theta_{i}(u_{0}) \phi^{i} (y(u_{0}))
    \frac{ (z_{\mu} - y_{\mu}(u_{0})) {\ddot y}_{\nu}(u_{0}) +
   (z_{\nu} - y_{\nu}(u_{0})) {\ddot y}_{\mu}(u_{0}) }
   { |z-y(u_{0})|^{3}  } W[C ] \nonumber \\
   & & + \frac{1}{12 \pi}  {\dot \theta}_{i}(u_{0}) \phi^{i} (y(u_{0}))
   \frac{ [ (z_{\mu} - y_{\mu}(u_{0})) {\dot 
   y}_{\nu}(u_{0}) + (z_{\nu} - y_{\nu} (u_{0})) {\dot y}_{\mu}(u_{0})
    ]}{ |z-y(u_{0})|^{3} } W[C ] \nonumber \\
   & & + \frac{1}{12 \pi} \theta_{i}(u_{0}) ( {\dot y}_{\alpha} (u_{0})
   \partial^{\alpha} \phi^{i} (y(u_{0}))) \nonumber \\
   & & \hspace{5mm} \times  \frac{ [ (z_{\mu} - y_{\mu}(u_{0})) {\dot 
   y}_{\nu}(u_{0}) + (z_{\nu} - y_{\nu} (u_{0})) {\dot y}_{\mu}(u_{0})
    ]  } { |z-y(u_{0})|^{3} } W[C ]. \label{AZMBsca2}
  \end{eqnarray}
 The other  term that requires a non-trivial computation is
  \begin{eqnarray}
   & & \frac{1}{3} \left[ \oint_{C} du  \theta^{i}(u)
 (\partial^{\alpha} \phi_{i}(y(u))) (z_{\alpha} - y_{\alpha}(u)) 
 ( \partial_{\mu}  \partial_{\nu} D(z-y(u)) ) \right] W[C ] \nonumber \\
  &=& \frac{1}{3} \int^{+\infty}_{-\infty} du \left[ \theta^{i}(u_{0})
 \partial^{\alpha} \phi_{i} (y(u_{0})) + 
 \left( \frac{d}{du}(\theta^{i}(u_{0}) \partial^{\alpha} \phi_{i}
 (y(u_{0})) ) \right)  (u-u_{0}) + \cdots  \right] \nonumber \\   
  & & \hspace{-5mm} \times
 \left[ \frac{(z_{\alpha} - y_{\alpha}(u_{0})) g_{\mu \nu}}{2 \pi^{2}
 [ (z_{\mu} - 
  y_{\mu}(u_{0}))^{2} + (u-u_{0})^{2} ]^{2} } - \frac{4 (z_{\mu} -
 y_{\mu}(u_{0})) (z_{\nu} - y_{\nu}(u_{0})) (z_{\alpha} -
 y_{\alpha}(u_{0})) }{2 \pi^{2}  [ (z_{\mu} - y_{\mu}(u_{0}))^{2} +
 (u-u_{0})^{2} ]^{3}} \right. \nonumber \\
  & &  +  \frac{4}{2 \pi^{2}  [ (z_{\mu} - y_{\mu}(u_{0}))^{2} +
 (u-u_{0})^{2} ]^{3} } (u-u_{0}) [  (z_{\mu} - y_{\mu}(u_{0})) (z_{\nu} -
 y_{\nu}(u_{0})) {\dot y}_{\alpha}(u_{0}) \nonumber \\
  & & \hspace{5mm} + (z_{\nu} - y_{\nu}(u_{0}))
 (z_{\alpha} - y_{\alpha}(u_{0})) {\dot y}_{\mu}(u_{0}) \nonumber \\
  & & \hspace{5mm}  + (z_{\alpha} -
 y_{\alpha}(u_{0})) (z_{\mu} - y_{\mu}(u_{0})) {\dot y}_{\nu}(u_{0}) +
 \cdots   ]  \Biggr] W[C ] \nonumber \\ 
 &=& \frac{1}{12 \pi |z-y(u_{0})|^{3} } g_{\mu \nu} \theta_{i} (u_{0}) 
   (z_{\alpha} - y_{\alpha} (u_{0}) ) (\partial^{\alpha} \phi^{i}
   ( y (u_{0})) ) W[C ] \nonumber \\
   & & - \frac{1}{12 \pi|z-y(u_{0})|^{5} } \theta_{i}(u_{0}) (z_{\alpha} 
   - y_{\alpha} (u_{0}) ) (\partial^{\alpha} \phi^{i} (y(u_{0})) )
   (z_{\mu} - y_{\mu}(u_{0}) ) (z_{\nu} - y_{\nu}(u_{0}) ) W[C ]
   \nonumber  \\
   & & - \frac{1}{12 \pi |z-y(u_{0})|^{3} }
   \theta_{i} (u_{0}) (z_{\alpha} - y_{\alpha}(u_{0}))
   (\partial^{\alpha} \phi^{i} (y(u_{0})) ) {\dot y}_{\mu} (u_{0})
   {\dot y}_{\nu} (u_{0}) W[C ] \nonumber \\
   & & - \frac{\theta_{i} (u_{0}) {\dot y}_{\alpha} (u_{0})
   (\partial^{\alpha} 
   \phi^{i} (y(u_{0})) )}{12 \pi |z-y(u_{0})|^{3}} \nonumber \\
  & & \hspace{5mm} \times [ (z_{\mu} - y_{\mu}(u_{0}) ) {\dot
   y}_{\nu}(u_{0}) + (z_{\nu} - y_{\nu}(u_{0}) ) {\dot y}_{\mu}(u_{0}) 
    ] W[C]. \label{AZMBsca3}
  \end{eqnarray}
 
 Summing the results (\ref{AZMBsca1}) $-$
 (\ref{AZMBsca3}),  we reach the conclusion (\ref{AZ22scares}).

\end{document}